\documentclass{emulateapj}
\newcommand{\nab}{\mbox{\boldmath $\nabla$} {}}
\newcommand{\BB}{\mbox{\boldmath $B$} {}}
\newcommand{\FF}{\mbox{\boldmath $F$} {}}
\newcommand{\JJ}{\mbox{\boldmath $J$} {}}
\newcommand{\VV}{\mbox{\boldmath$V$}}
\newcommand{\kk}{\mbox{\boldmath $k$} {}}
\newcommand{\vv}{\mbox{\boldmath $v$} {}}
\newcommand{\xx}{\mbox{\boldmath $x$} {}}
\newcommand{\yy}{\mbox{\boldmath $y$} {}}
\newcommand{\zz}{\mbox{\boldmath $z$} {}}
\newcommand{\rr}{\mbox{\boldmath $r$} {}}
\newcommand{\bb}{\mbox{\boldmath $\beta$} {}}
\newcommand{\oo}{\mbox{\boldmath $\omega$} {}}

\newcommand{\mach}{\mathcal{M}}

\shorttitle{DYNAMICAL FRICTION }
\shortauthors{S{\'A}NCHEZ-SALCEDO }

\begin{document}
\title{Dynamical Friction in a Gaseous Medium with a Large-Scale
Magnetic Field}
\author{
  F.~J.~S{\'a}nchez-Salcedo\altaffilmark{1}}
\altaffiltext{1}{Instituto de Astronom\'ia, Universidad Nacional Aut\'onoma 
de M\'exico, Ciudad Universitaria, 04510 Mexico City, Mexico; 
jsanchez@astroscu.unam.mx}

\begin{abstract}
The dynamical friction force experienced by a massive gravitating body moving 
through
a gaseous medium is modified by sufficiently strong large-scale magnetic fields.
Using linear perturbation theory,
we calculate the structure of the wake generated by, and the gravitational drag force on,
a body traveling in a
straight-line trajectory in a uniformly magnetized medium.
The functional form of the drag force as a function of the Mach number ($\equiv V_{0}/c_{s}$,
where $V_{0}$ is the velocity of the body and $c_{s}$ the sound speed) depends on
the strength of the magnetic field and on the angle between the velocity of the 
perturber and the direction of the magnetic field. 
In particular, the peak value of the
drag force is not near Mach number $\sim 1$ for a perturber moving
in a sufficiently magnetized medium. 
As a rule of thumb, we may state that for supersonic motion,
magnetic fields act to suppress dynamical friction; 
for subsonic motion, magnetic fields
tend to enhance dynamical friction. For perturbers moving
along the magnetic field lines, the drag force at some subsonic Mach numbers may
be stronger than it is at supersonic velocities. We also mention the relevance of our
findings to black hole coalescence in galactic nuclei.

\end{abstract}



\section{Introduction}
\label{sec:introduction}

An  object moving in a background medium induces a gravitational wake.
The asymmetry of the mass density distribution upstream and downstream
from the perturber produces a drag on the body, which is often referred to
as gravitational drag or dynamical friction (DF) force. 
A body in orbital motion may undergo a radial decay of its orbit
due to the loss of angular momentum by the negative torque
caused by DF drag.
Chandrasekhar (1943) derived 
the dynamical friction on a massive particle passing through a homogeneous
and isotropic background of light stars. His formula is applied to estimate the
merger timescale of satellite systems or to study the accretion history 
of galaxies.  Bondi \& Hoyle (1944) considered the problem of the mass accretion by
a point mass $M$ travelling at velocity $V_{0}$ in a collisional homogeneous medium of sound speed
$c_{s}$ in the limit
where the perturber moves at supersonic
velocities relative to the ambient gas (i.e.~high Mach numbers). 
If the perturber is an accretor, streamlines with small impact parameter may
become bound because of energy dissipation in shocks, and can be accreted to the 
perturber.
Hence, the force on the perturber consists of two parts; the gravitational drag
and the momentum accretion force. The latter contribution may be 
decelerating or accelerating (Ruffert 1996).
If the size of the perturber is larger than the Bondi-Hoyle accretion radius defined
as $R_{BH}\equiv GM/[c_{s}^{2}(1+\mach^{2})]$ with $\mach=V_{0}/c_{s}$, 
the density and 
velocity structure of the wake, at any Mach number,
can be inferred
analytically in linear theory because the body produces a small perturbation 
in the ambient gaseous medium at any location. The gravitational drag is inferred
as the gravitational attraction between the perturber and its own wake (e.g., Dokuchaev 1964; 
Rephaeli \& Salpeter 1980; Just \& Kegel 1990; Ostriker 1999; Kim \& Kim 2007;
S\'anchez-Salcedo 2009; Namouni 2010). 

The studies of the gravitational drag in gaseous media have enjoyed
widespread theoretical application, ranging from protoplanets to galaxy 
clusters.
It seems to play a significant role 
in the growth of planetesimals
(Hornung, Pellat \& Barge 1985; Stewart \& Wetherill 1988), the 
eccentricity excitation of planetary embryo orbits (Ida 1990;
Namouni et al.~1996), the orbital decay of common-envelope binary
stars (e.g., Taam \& Sandquist 2000; Nordhaus \& Blackman 2006; 
Ricker \& Taam 2008; Maxted et al. 2009; Stahler 2010),
the evolution of the orbits of  planets around the
more massive stars (Villaver \& Livio 2009),
the evolution of low-mass condensations in the cores of molecular clouds
(Nejad-Asghar 2010),
the mass segregation of massive stars in young clusters  
embedded in dense molecular cores (Chavarr\'{\i}a et al.~2010),
the orbital decay of kpc-sized giant clumps in galaxies at high redshift
\citep{imm04,bou07},
or the heating of intracluster gas by supersonic galaxies (El-Zant
et al.~2004; Kim et al.~2005; Kim 2007; Conroy \& Ostriker 2008).  
Special work has been devoted to understand the role of
gaseous DF in the orbital decay of stars and supermassive black holes
as a result of hydrodynamic interactions with an accretion flow 
in galactic nuclei (Narayan 2000).
Mergers of supermassive black hole binaries may be accelerated on sub-parsec
scales by angular momentum loss to surrounding gas (Armitage \& Natarajan 2005).
In particular, 
gaseous DF expedites the growth of SMBH by mergers in colliding galaxies
(Escala et al.~2004, 2005; Dotti et al.~2006;
Mayer et al.~2007; Tanaka \& Haiman 2009; Colpi \& Dotti 2011).

Less developed is the corresponding theory of DF in a magnetized gaseous medium.
As far as we know, the analytic estimate of the gravitational drag 
for a body moving on a rectilinear trajectory {\it parallel} to
the uniform unperturbed magnetic field lines by 
Dokuchaev (1964) is the only work in this area.
He concluded that the DF force on a supersonic body is reduced 
by a factor that depends on
the ratio between the Alfv\'en speed and the sound speed.
Since large-scale magnetic fields are ubiquitous in many astronomical systems
such as molecular clouds (Tamura \& Sato 1989; Goodman \& Heiles 1994; 
Matthews \& Wilson 2002; Heiles \& Crutcher 2005) or galactic nuclei, it is
important to understand how the DF force is affected by the presence of 
ordered large-scale magnetic fields.
In fact, young stellar systems and low-mass condensations orbiting in the potential
of their birth clusters can interact with the surrounding dense and magnetized
molecular interstellar medium during the dispersal of the cluster's gas.
In the Galactic center, structures associated with ordered magnetic fields,
called arches and threads, are detected in radio continuum maps
(Yusef-Zadeh et al.~1984). 
The magnetic field configuration of the Galactic center has been viewed
as poloidal in the diffuse, interstellar (intercloud) medium and
approximately parallel to the Galactic plane only in the dense
molecular clouds (Nishiyama et al. 2010). On the scale of $400$ pc,
fields of $100\mu$G have been reported (Chuss et al.~2003; Crocker et al.~2010).

The importance of gaseous DF in the evolution and coalescence of a massive black hole binary 
is motivated by both observational and theoretical work that indicate
the presence of large amounts of gas in the central region of merging galaxies.
During the merger of galaxies, the inflow of gas material towards the
galactic center driven by tidal torques associated with bar instabilities
and shocks will sweep up and amplify the magnetic field in the central region
(Callegari et al.~2009; Guedes et al.~2011).
Observations of gas-rich interacting galaxies such as the
ultraluminous infrared galaxies (ULIRGs) show that their central regions contain
massive and dense clouds of molecular and atomic gas (Sanders \& Mirabel 1996).
ULIRGs are natural locations to expect very strong magnetic fields 
(Thompson et al.~2006; Robishaw et al.~2008; Thompson et al.~2009).

In this paper we will study the DF in a gaseous medium on a body moving 
on rectilinear orbit in a homogeneous, uniform magnetized cloud.
This is the simplest idealized extension of the unmagnetized case and is
the first step in understanding the role of ordered magnetic fields.
Previous works have shown that, 
although the formulae of the gaseous drag force in a {\it unmagnetized} 
gas medium, were derived for rectilinear orbits in homogeneous and
infinite media (Dokuchaev 1964; Rephaeli \& Salpeter 1980; Just  \& Kegel 1990;
Ostriker 1999; S\'anchez-Salcedo \& Brandenburg 1999; Kim \& Kim 2009; 
Namouni 2010; Lee \& Stahler 2011; Cant\'o et al.~2011), 
simple `local' extensions have been proven very successful
in more realistic situations, e.g.~smoothly decaying density backgrounds or 
when the perturber is moving on a circular orbit (S\'anchez-Salcedo \& 
Brandenburg 2001; Kim \& Kim 2007; Kim et al.~2008; Kim 2010).
As a useful starting point for understanding the role of a large-scale
magnetic field, we also consider that the unperturbed medium is homogeneous and 
uniformly magnetized.
A discussion on the DF in other initial force-free configurations will be
given in a separate paper.
Even in the simple case of a uniformly magnetized medium, the magnetic field
produces qualitatively new phenomena.

The paper is organized as follows. In \S\ref{sec:basic}, 
we discuss the basic concepts
on the ideal problem of a particle traveling at constant speed through 
a uniform gas, both in the purely hydrodynamic case and when the
plasma is magnetized.
In \S\ref{sec:axisymmetric_steady}, we outline the linear 
derivation for calculating the steady-state density wake generated by an
 extended body moving along the magnetic fields, give an analytical
solution of the problem and compare it with previous work.
The time-dependent linear perturbation theory is presented in \S\ref{sec:time_dependent}.
\S\ref{sec:results} describes the structure of the resulting wake and evaluate the
DF force as a function of Mach number,
for different angles between the direction of the perturber's velocity and
the direction of the magnetic field.
In \S\ref{sec:summary}, we summarize our results and discuss their implications.

\section{Dynamical friction in gaseous media: Basic formulae}
\label{sec:basic}
\subsection{Unmagnetized medium}
\label{sec:unmagnetizedmedium}
Under assumption of a steady state, Dokuchaev (1964), Ruderman \& Spiegel (1971)
and Rephaeli \& Salpeter (1980) derived the drag
force on a point mass  $M$ moving at velocity $V_{0}$ on a straight-line
trajectory through a uniform medium with unperturbed density $\rho_{0}$ 
and sound speed $c_{s}$.
For subsonic perturbers (${\mathcal{M}}\equiv V_{0}/c_{s}<1$, 
where ${\mathcal{M}}$ is
the Mach number), these authors found that the drag force 
 is zero because of the front-back symmetry of the density distribution
about the perturber. For the steady-state supersonic case, the drag force takes the form
\begin{equation}
F_{DF}=\frac{4\pi \rho_{0} (GM)^{2}}{V_{0}^{2}}\ln\Lambda,
\label{eq:rephaeli}
\end{equation}
where $\Lambda=r_{\rm max}/r_{\rm min}$, being $r_{\rm max}$ and $r_{\rm min}$ the maximum and minimum radii
of the effective gravitational interaction of a perturber with the gas.
For extended perturbers, $r_{\rm min}$ is its characteristic size, whereas for
pointlike perturbers $r_{\rm min}$ is of the order of the Bondi-Hoyle 
radius $R_{BH}$ (Cant\'o et al.~2011), 
as defined in \S \ref{sec:introduction}.

Using a time-dependent analysis in the unmagnetized case, Ostriker (1999) found
that (1) the force is not zero at the subsonic regime because,
although a subsonic perturber generates a density distribution with
contours of constant density corresponding to ellipsoids, 
there are always cut-off ones
within the sonic sphere that exert a gravitational drag, and (2) $r_{\rm max}$ increases with
time in the supersonic case.
More specifically, she found that the Coulomb logarithm is given by:
\begin{equation}
\ln\Lambda=\frac{1}{2}\ln \left(\frac{1+\mach}{1-\mach}\right)
-\mach,
\label{eq:ostriker1}
\end{equation}
for $\mach<1$ and $t>r_{\rm min}/(c_{s}-V_{0})$, and
\begin{equation}
\ln\Lambda=\frac{1}{2}\ln \left(1-\mach^{-2}\right)+
\ln\left(\frac{V_{0}t}{r_{\rm min}}\right),
\label{eq:ostriker2}
\end{equation}
for $\mach>1$ and $t>r_{\rm min}/(V_{0}-c_{s})$. The perturber is assumed to be 
formed at $t=0$. 
The transition between the subsonic to the supersonic regime is smooth
without any divergence at a Mach number of unity (see Fig.~3 in Ostriker 1999). 
S\'anchez-Salcedo \& Brandenburg (1999) tested numerically
that Ostriker's formula is very accurate for non-accreting extended perturbers. 
In many astrophysical situations, one needs to assign a softening radius to the
gravitational potential which in turn determines $r_{\rm min}$ without any 
ambiguity.  
For a body described with a Plummer model with core radius $R_{\rm soft}\gg R_{BH}$,
S\'anchez-Salcedo \& Brandenburg (1999) found that $r_{\rm min}\simeq 
2.25R_{\rm soft}$.

For point-mass accretors, the friction force has been derived
by Lee \& Stahler (2011) in the subsonic regime and by 
Cant\'o et al.~(2011) in the hypersonic limit.

\subsection{Magnetized medium}
\label{sec:magnetizedmedium}
The presence of a small-scale magnetic field tangled at scales below $r_{\rm min}$ will change
the speed of sound. For isotropic compression of a random magnetic field, 
the effective sound speed is $(c_{s}^{2}+\frac{2}{3}c_{a}^{2})^{1/2}$ (e.g., Zweibel 2002),
where $c_{a}$ is the Alfv\'en speed of the random small-scale component
of the magnetic field. Therefore, in order to include the effect of a small-scale magnetic field,
one has to replace the sound speed by the effective sound speed in the
definition of $\mach$ in Eqs.~(\ref{eq:ostriker1})
and (\ref{eq:ostriker2}).

The extension of the drag force formulae is by no means
straightforward if the gaseous medium is permeated by a regular magnetic
field. Dokuchaev (1964) derived the gravitational drag force in the steady state for
perturbers moving along the lines of the unperturbed magnetic field.
He found that the DF drag is 
\begin{equation}
F_{DF}=\left(1-\frac{c_{A}^{2}}{V_{0}^{2}}\right)\frac{4\pi \rho_{0} 
(GM)^{2}}{V_{0}^{2}}\ln\Lambda,
\label{eq:dokuchaev_drag}
\end{equation}
at $V_{0}>(c_{s}^{2}+c_{A}^{2})^{1/2}$, where $c_{A}$ is the Alfv\'en speed
of the regular magnetic field, and it is zero for 
$V_{0}<(c_{s}^{2}+c_{A}^{2})^{1/2}$. 
By comparing Eqs.~(\ref{eq:rephaeli})
and (\ref{eq:dokuchaev_drag}), we see that the drag in a uniform magnetized background 
is never larger than in the unmagnetized case. 
According to Dokuchaev (1964), the gravitational drag on a body with 
velocity $V_{0}$ in a uniformly magnetized medium is equal
to the drag on a body with velocity $V_{0}/(1-c_{A}^{2}/V_{0}^{2})^{1/2}$
in a unmagnetized medium.
Therefore, if one naively uses the nonmagnetic formulae (\ref{eq:rephaeli})-(\ref{eq:ostriker2})
by replacing the sound speed $c_{s}$ by the magnetosonic speed 
$(c_{s}^{2}+c_{A}^{2})^{1/2}$
would yield to wrong results. 
In the next Section, we will show, however, that the paper of Dokuchaev
(1964) contains an error and it is not true that the drag force in
the magnetized medium case is always smaller than in the unmagnetized case.

As Ostriker (1999) demonstrated in the field-free case, the steady state 
result found by Dokuchaev that the net force is zero
at $V_{0}<(c_{s}^{2}+c_{A}^{2})^{1/2}$, because of the front-back symmetry of the density 
perturbation in the medium, may be misleading. It is also unclear how
$\ln\Lambda$ varies in time for the magnetized supersonic case. 
Moreover, it left unexplored how the drag force
depends on the angle between the velocity of the perturber and the direction of the
magnetic field. Before addressing these questions, however, it is still worthwhile
finding analytical solutions for the perturbed steady
density and the resulting drag force in the simplest scenario in which
the velocity of the perturber and the magnetic field are parallel.
Such a exact treatment will allow us to gain insight into more complicated
situations. This will be done in the next Section.

\section{Axisymmetric case: Velocity of the perturber parallel to the direction
of the magnetic field}
\label{sec:axisymmetric_steady}
We consider a gravitational perturber moving 
on a straight-line at constant velocity 
in a medium with unperturbed density $\rho_{0}$ 
and thermal sound speed $c_{s}$.
In the absence of magnetic fields, the linearized equations of motion can be reduced
to a nonhomogeneous wave equation for the relative perturbation $(\rho-\rho_{0})/\rho_{0}$
(e.g., Ruderman \& Spiegel 1971; Ostriker 1999). Once a uniform magnetic field, $\BB_{0}$,  parallel
to the direction of perturber's velocity  is included, Dokuchaev (1964) showed 
that the relative perturbation
obeys an equation of fourth order in $t$ and solved it using a double Fourier-Hankel
transformation. As it will become clear later, we prefer to describe the evolution of
the system through wave-equations because it facilitates the physical interpretation of the
problem and because the contact with the analysis of Ostriker (1999) is easier. In addition,
the extension of the equations for a case where the angle between $\BB_{0}$ 
and the velocity of the perturber  is arbitrary, 
becomes straightforward in our approach. 

\subsection{Perturbed density distribution}
\label{sec:perturbed_density}
We study first the completely
steady flow created by a mass on a constant-speed trajectory parallel
to the lines of the unperturbed magnetic field $\BB_{0}=B_{0}
\hat{\zz}$.
To do this, consider a particle at the origin of our coordinate system, surrounded by a gas
whose velocity far from the particle is $\VV_{0}=-V_{0}\hat{\zz}$, with
$V_{0}>0$. We will further assume that the gas evolves under
flux-freezing conditions.
Our analysis begins with the linearized MHD equations to describe
the medium's response to the perturber's presence $\rho=\rho_{0}+\rho'$, 
$\VV=\VV_{0}+
\vv'$ and $\BB=\BB_{0}+\BB'$,  in a stationary sate ($\partial/\partial t=0$)
\begin{equation}
\rho_{0} \nab\cdot \vv'+\VV_{0}\cdot\nab\rho'=0,
\label{eq:steady_continuity}
\end{equation}
\begin{equation}
(\VV_{0}\cdot \nab)\vv'=
-\frac{c_{s}^{2}\nab\rho'}{\rho_{0}}
-\nab\Phi + \frac{1}{4\pi\rho_{0}} (\nab\times \BB')\times
\BB_{0},
\label{eq:steady_motion}
\end{equation}
\begin{equation}
\nab\times (\VV_{0}\times \BB')+\nab\times
(\vv'\times \BB_{0})=0,
\label{eq:steady_induction}
\end{equation}
\begin{equation}
\nab\cdot\BB'=0,
\end{equation}
where $\Phi$ is the gravitational potential created by the perturber. The Poisson equation
links the potential with the density profile of the perturber $\rho_{p}$:
\begin{equation}
\nabla^{2}\Phi=4\pi G \rho_{p}.
\label{eq:poisson}
\end{equation}

The Lorentz force, which provides the magnetic back-reaction on
the flow pattern, is given by 
\begin{equation}
(\nab\times \BB')\times \BB_{0}
=B_{0}\left( \left[\frac{\partial B_{x}'}{\partial z}-
\frac{\partial B_{z}'}{\partial x} \right]\hat{\xx}-
\left[\frac{\partial B_{z}'}{\partial y}-
\frac{\partial B_{y}'}{\partial z} \right]\hat{\yy}
\right).
\label{eq:lorentzforce}
\end{equation}
Hence, the divergence of the Lorentz force is
\begin{eqnarray}
&&\nonumber\nab\cdot\left[(\nab\times \BB')\times \BB_{0}\right]=
\\&&\nonumber=B_{0}\left(\frac{\partial^{2} B_{x}'}{\partial x \partial z}-
\frac{\partial^{2} B_{z}'}{\partial x^{2}}
-\frac{\partial^{2} B_{z}'}{\partial y^{2}}
+\frac{\partial^{2} B_{y}'}{\partial y \partial z}
\right)\\&&=-B_{0}\nabla^{2} B_{z}'.
\label{eq:div_Lorentz}
\end{eqnarray}
In the last equality we have used that $\nabla\cdot \BB'=0$.
By substituting equations (\ref{eq:steady_continuity}) and 
(\ref{eq:div_Lorentz}) 
in the divergence of the equation of motion\footnote{The curl of the equation of motion provides a relationship
between the vorticity $\oo$ and the current density $\JJ'$:
\begin{eqnarray}
-V_{0}\frac{\partial\oo}{\partial z} 
=\frac{B_{0}}{c \rho_{0}}\frac{\partial \JJ'}{\partial z}. 
\nonumber
\end{eqnarray}
In linear theory, the baroclinic term vanishes and the Lorentz term
is the only able to generate vorticity, even if
the gravitational force is irrotational.},
we have 
\begin{equation}
-\frac{V_{0}^{2}}{\rho_{0}}\frac{\partial^{2}\rho'}{\partial z^{2}}
=-\frac{c_{s}^{2}}{\rho_{0}}\nabla^2 \rho' -\nabla^{2} \Phi-
\frac{B_{0}}{4\pi \rho_{0}}\nabla^{2}B_{z}'.
\label{eq:ostriker_gen}
\end{equation}
By comparing the second and third terms in the right-hand side
of the equation above, we see that, formally,
the magnetic back-reaction term $\nabla^{2}B_{z}'$ is mathematically equivalent
to having an external potential term.
However, whilst $\Phi$ is known (Eq.~\ref{eq:poisson}),
$B_{z}'$ is coupled to the fluid motions through the flux-freezing equation
(\ref{eq:steady_induction}).

Next, we need an independent equation for $B_{z}'$ to close the system.
This can be accomplished using the third component of the induction equation
(\ref{eq:steady_induction}), which has the form
\begin{equation}
B_{0}\left(\nab\cdot\vv'-\frac{\partial v_{z}'}{\partial z}\right)=
V_{0}\frac{\partial B_{z}'}{\partial z}.
\label{eq:induction99}
\end{equation}
Our strategy is to eliminate $\vv'$ in Equation (\ref{eq:induction99}).
The third component of the equation of motion (\ref{eq:steady_motion})
can be written as
\begin{equation}
-V_{0}\frac{\partial v_{z}'}{\partial z}=-\frac{c_{s}^{2}}{\rho_{0}}
\frac{\partial \rho'}{\partial z}-\frac{\partial \Phi}{\partial z}.
\label{eq:steady_third}
\end{equation}
This equation does not depend explicitly on the frozen-in magnetic field
because the $z$-component of the Lorentz force vanishes in the linear
approximation (see Eq.~\ref{eq:lorentzforce}). 
Substituting Eqs. (\ref{eq:steady_continuity}) and (\ref{eq:steady_third}) 
in Eq. (\ref{eq:induction99}), we obtain the desired equation
\begin{equation}
B_{0}\left(\frac{1}{\rho_{0}} \frac{\partial\rho'}{\partial z}
-\frac{1}{\rho_{0}{\mathcal{M}}^{2}} \frac{\partial\rho'}{\partial z}
-\frac{1}{V_{0}^{2}}\frac{\partial\Phi}{\partial z}\right)
=\frac{\partial B_{z}'}{\partial z},
\label{eq:steady_second}
\end{equation}
where we recall that ${\mathcal{M}}\equiv V_{0}/c_{s}$ is the (sonic)
Mach number.
Once again, the magnetic term $\partial B'_{z}/\partial z$
is formally identical to $\partial \Phi/\partial z$, but some caution
should be used when interpreting it;  the $z$-component of
the Lorentz force is not $\partial B_{z}'/\partial z$ but zero.

Equations (\ref{eq:ostriker_gen}) and (\ref{eq:steady_second}) 
constitute a system of two coupled linear differential
equations for $\rho'$ and $B_{z}'$ which may be solved once we have chosen
suitable boundary conditions.
Defining the dimensionless perturbations $\alpha\equiv \rho'/\rho_{0}$
and $\bb\equiv \BB'/B_{0}$, the equations to solve are:
\begin{equation}
{\mathcal{M}}^{2}\frac{\partial^{2} \alpha}{\partial z^{2}}
=\nabla^{2}\alpha +\frac{1}{c_{s}^{2}}\nabla^{2}\Phi
+\Upsilon^{2}\nabla^{2}\beta_{z},
\label{eq:steady_ostriker_gen}
\end{equation}
\begin{equation}
({\mathcal{M}}^{2}-1)\frac{\partial \alpha}{\partial z}
-\frac{1}{c_{s}^{2}}\frac{\partial \Phi}{\partial z}
={\mathcal{M}}^{2}\frac{\partial \beta_{z}}{\partial z},
\label{eq:goesFourier2}
\end{equation}
where $\Upsilon\equiv c_{A}/c_{s}$ and $c_{A}$ the Alfv\'en speed in the unperturbed medium.
In the limit of vanishing magnetic field, $\beta_{z}=\Upsilon=0$ and 
Equation (\ref{eq:steady_ostriker_gen}) 
reduces to that of the wake of a body in a unnmagnetized medium 
(e.g., S\'anchez-Salcedo 2009).

For a point-like perturber of mass $M$, an analytical solution can be 
derived for the 
density enhancement, velocity and magnetic fields in the wake.
In order to calculate the dynamical friction force exerted on the body, 
we only need the gas density enhancement in the wake, which is 
derived in Appendix \ref{sec:appendix1} and is given by
 \begin{equation}
\alpha(R,z)=\frac{\lambda(1-\eta)GM}{\xi c_{s}^{2}}
\frac{1}{\sqrt{z^{2}+R^{2}\gamma^{2}}},
\label{eq:alpha_axisymmetric}
\end{equation}
where $R=\sqrt{x^2+y^2}$ is the cylindrical radius and
\begin{equation}
\eta=(c_{A}/V_{0})^{2}=(\Upsilon/{\mathcal{M}})^{2},
\label{eq:eta_def}
\end{equation}
\begin{equation}
\xi= 1+(1-\mach^{-2})\Upsilon^{2}=1-\eta+\Upsilon^{2},
\label{eq:xi_def}
\end{equation}
\begin{equation}
\gamma^{2}=1-\frac{{\mathcal{M}}^{2}}{\xi},
\label{eq:gamma_def}
\end{equation}
and
\begin{equation}
                        \lambda = \left\{ \begin{array}{ll}
         1 & \mbox{if ${\mathcal{M}}<{\mathcal{M}}_{\rm crit}$; }\\
         2 & \mbox{if ${\mathcal{M}}_{\rm crit}<{\mathcal{M}}<{\rm min}(1,\Upsilon)$ and 
              $z/R>|\gamma|$};\\
         1   & \mbox{if ${\rm min}(1,\Upsilon)<{\mathcal{M}}<{\rm max}(1,\Upsilon)$};\\
         2 & \mbox{if ${\mathcal{M}}>{\rm max}(1,\Upsilon)$ and $z/R<-|\gamma|$};\\
         0 & \mbox{otherwise.} \end{array} \right. 
\label{eq:lambda_complete}
\end{equation}
Here, the critical Mach number is defined as
\begin{equation}
{\mathcal{M}}_{\rm crit}\equiv \left(1+\Upsilon^{-2}\right)^{-1/2}.
\end{equation}
Because of the linear-theory assumption, Equation (\ref{eq:alpha_axisymmetric})
is properly valid only for $(z^{2}+\gamma^{2}R^{2})^{1/2}
\gg (1-\eta)GM/(\xi c_{s}^{2})$.
The nonmagnetic steady-state solution for density in the wake past a 
gravitating body is recovered when $\Upsilon=0$. 

For clarity, it is convenient to distinguish 
four intervals depending on the value of the Mach number of the body:
${\mathcal{M}}<{\mathcal{M}}_{\rm crit}$ (case or interval I),
${\mathcal{M}}_{\rm crit}<{\mathcal{M}}<{\rm min}(1,\Upsilon)$ (case or
interval II),
${\rm min}(1,\Upsilon)<{\mathcal{M}}<{\rm max}(1,\Upsilon)$ (case III)
and ${\mathcal{M}}>{\rm max}(1,\Upsilon)$ (interval IV). 
$\gamma^{2}$ is a positive number in cases I and III, whereas it
is negative in cases II and IV.
In the latter cases where $\gamma^{2}<0$, the density perturbation
$\alpha$ vanishes at some spatial
locations. In case IV, for instance, $\alpha$ 
outside the cone defined by the surface $z=-|\gamma|R$ is actually zero.
Turning to Eq.~(\ref{eq:steady_second}), we see that the magnetic 
perturbation $B_{z}'$
in these regions does not vanish but obeys the following relation,
$\partial B_{z}'/\partial z=-(B_{0}/V_{0}^{2})\partial \Phi/\partial z$.
Now, from Eq.~(\ref{eq:steady_third}), 
the axial component of the velocity satisfies
a similar equation, $\partial v_{z}'/\partial z= (1/V_{0})\partial \Phi/
\partial z$. Using the fact that $\nab\cdot \vv' =0$ in regions of
constant density (see Eq.~\ref{eq:steady_continuity}), 
it is simple to show that $\BB'$ is parallel to
$\vv'$ in regions where $\alpha=0$, and thus the magnetic 
configuration is force-free in these zones.

Once $\rho/\rho_{0}$ is known, the gravitational drag can be computed;
this will be done in Section \ref{sec:dragforce_axi}. 
Nevertheless, in order to gain
more insight into the physics of the wake, we will describe the
morphology and structure of the steady-state wake in the next Section.

\subsection{Physical interpretation}
\label{sec:physical_interpretation}
Consider first subsonic perturbers.
In the limit $̣{\mathcal{M}}\rightarrow 0$, we have $\gamma\rightarrow 1$,
$\xi\rightarrow -\Upsilon^{2}/{\mathcal{M}}^{2}$, $(1-\eta)\rightarrow
-\Upsilon^{2}/{\mathcal{M}}^{2}$ and $\lambda=1$. Therefore,
the density enhancement is $GM/(c_{s}^{2}r)$, which is 
$\Upsilon$-independent, and corresponds to the linearized solution of the
hydrostatic envelope, $\rho/\rho_{0}=\exp \left[GM/(c_{s}^{2}r)\right]$,
around a stationary perturber (e.g., Ostriker 1999).
In this case, the magnetic field lines remain straight and the
whole magnetic configuration is force-free.

The surfaces of constant density for subsonic perturbers may be
either ellipsoids or hyperbolae, depending on the Mach number.
At ${\mathcal{M}}<{\mathcal{M}}_{\rm crit}$ it holds that $\gamma^{2}>1$ and
the isodensity surfaces are ellipsoids 
{\it elongated} along the trajectory of the perturber with
eccentricity $e={\mathcal{M}}/\sqrt{|\xi|}$. 
This is in sharp contrast to what happens without any magnetic field where
the ellipses are elongated along $R$ for perturbers at {\it any
subsonic Mach number} ($\gamma^{2}<1$). Therefore, the existence
of ellipsoidal density isocontours elongated along $z$ 
is a clear signature of curved magnetic fields.

In the {\it nonmagnetic subsonic case}, $v_{R}'<0$ at $z>0$ and $v_{z}'>0$
at any $z$, signifying that the incoming fluid is veering towards the
perturber, but then turning away again once it passes the body. 
The result of $v_{z}'>0$ implies that the gas is being dragged
by the gravitational pushing of the body. In order to understand
the morphology of the wake in a magnetized medium, in the following we 
calculate $v_{R}'$ and $v_{z}'$ for
a perturber with Mach number in the interval I. 

From Eq.~(\ref{eq:steady_third}) and using the result for $\alpha$ in 
Eq.~(\ref{eq:alpha_axisymmetric}), we find that
\begin{equation}
V_{0}\frac{\partial v_{z}'}{\partial z}=-\frac{\lambda (1-\eta)}{\xi }
\frac{GMz}{(z^{2}+R^{2}\gamma^{2})^{3/2}}
+\frac{GMz}{(z^{2}+R^{2})^{3/2}},
\end{equation} 
which leads to
\begin{equation}
v_{z}'=-\frac{GM}{V_{0}}\left(\frac{1}{(z^{2}+R^{2})^{1/2}}-
\frac{\lambda (1-\eta)}{\xi(z^2+R^2 \gamma^2)^{1/2}}\right).
\end{equation}
It is simple to show that $v_{z}'>0$ in case I, regardless the value
of $\Upsilon$.

The radial component of the velocity can be found 
using Eq.~(\ref{eq:steady_continuity}), 
\begin{equation}
\frac{1}{R}\frac{\partial Rv_{R}'}{\partial R}=
-\frac{\partial v_{z}'}{\partial z}+V_{0}\frac{\partial \alpha}{\partial z}
=-\frac{V_{0}}{{\mathcal{M}}^{2}}\left((1-{\mathcal{M}}^{2})\frac{\partial \alpha}{\partial z}
+\frac{1}{c_{s}^{2}}\frac{\partial \Phi}{\partial z}\right).
\end{equation}
Since we already know $\alpha$ and $\Phi$, this equation can 
be solved to obtain the radial velocity:
\begin{equation}
v_{R}'=\frac{GMz}{V_{0}R}\left(\frac{1}{(z^{2}+R^{2})^{1/2}}-
\frac{\lambda (1-\mach^2)(1-\eta)}
{\gamma^{2}\xi(z^2+R^2 \gamma^2)^{1/2}}\right).
\end{equation}
From the equation above, it follows that, in case I, $v_{R}'>0$ at 
the head of the perturber, regardless the magnetic field strength. 
Thus, a parcel of fluid in the upstream region circulates
around the perturber, reaching its maximum $R$-value at $z=0$ and then
turning back again.
Since frozen-in magnetic field lines are dragged by the gas, 
the upstream magnetic field lines are decompressed radially, 
resulting in arched magnetic field lines with negative $B_{z}'$-values.
In fact, integration of Eq.~(\ref{eq:steady_second}), gives
\begin{equation}
B_{z}'=\frac{GMB_{0}}{V_{0}^{2}}\left(\frac{1}{(z^{2}+R^{2})^{1/2}}
-\frac{\lambda (1-\mach^{2})(1-\eta)}{\xi(z^{2}+R^{2}\gamma^{2})^{1/2}}\right).
\end{equation}
This clearly states that $B_{z}'\leq 0$ (here we assume $B_{0}>0$).
Thus, an anticorrelation between density and $B_{z}'$ arises. 

At Mach numbers close to ${\mathcal{M}}_{\rm crit}$, that is,
${\mathcal{M}}={\mathcal{M}}_{\rm crit}-\epsilon$ with $\epsilon$ a very
small positive number,
the density profile, at not extremely large $z$ distances, is
\begin{equation}
\alpha(R,z)\simeq \frac{\Upsilon^{3/2}GM/c_{s}^{2}}{\sqrt{2\epsilon}(1+\Upsilon^2)^{1/4}R}.
\end{equation}
Hence, the density enhancement is large and its $z$-gradient very 
small.

At {\it subsonic} Mach numbers in the interval
${\mathcal{M}}_{\rm crit}<{\mathcal{M}}<{\rm min}(1,\Upsilon)$, the surfaces
of constant density exhibit no front-back symmetry. The isodensity
contours correspond to hyperbolae in the $z-R$ plane, as occurs for
supersonic perturbers in the absence of magnetic fields,
but now the density perturbation is null in the rear Mach cone
and is non-vanishing in a modified Mach cone leading the perturber.
The physical reason is as follows.
The density perturbation $\alpha$ is non-positive at any location
because the streamlines diverge at the front cone ($\eta>1$ and $\xi>0$).
$\alpha$ at the edges of the cone is minus infinity for a point mass.
In the front cone, $\partial v_{z}'/\partial z >0$, meaning that the
flow in that region is being accelerated by the inward net pressure force.
Across the edge of the modified Mach cone, there is a rapid rise in
pressure and density, and the gas velocity quickly slows.
In fact, the causality criterion used in Appendix \ref{sec:appendix1}  
is tantamount to selecting the solution in which
a rapid flow is slowed in a short distance,
as occurs in shock waves. 
Indeed, we will find numerically further below
that the system adopts this solution (\S \ref{sec:axisymmetriccase}). 

In case III, the body moves at intermediate velocities, i.e.~either 
in the range $c_{s}<V_{0}<c_{A}$ (if
$\Upsilon>1$) or in the range $c_{A}<V_{0}<c_{s}$ (if $\Upsilon<1$).
It is only in this case
that $\gamma$ lies in the range $0<\gamma<1$ and, therefore, the ellipsoids
are flattened along $z$. Remind that in the unmagnetized background,
a subsonic perturber also produces elliptical density distributions 
flattened along $z$
(e.g., Ostriker 1999); the latter is a particular case of $c_{A}<V_{0}<c_{s}$.
For $c_{s}<V_{0}<c_{A}$ (which requires that $\Upsilon>1$), 
however, the density perturbation
is negative ($\alpha\leq 0$).  In this case, the thermal pressure
decreases towards the body.  
Therefore, ahead of the perturber the gas is accelerated along the 
$z$-direction by
the gravitational force plus the pressure gradient (it holds that
$\partial v_{z}'/\partial z>0$ at $z>0$).
A negative $\alpha$ is a consequence of the action of
the pressure gradient plus the reduction of the radial convergence
of the flow due to the presence of the ordered magnetic field that 
preferentially
allows motions along $z$, which lead to an accelerated flow falling towards
the perturber. In the radial direction, the magnetic field lines are
compressed at $z>0$ because $v_{R}<0$ upstream of the perturber.

From Eqs.~(\ref{eq:alpha_axisymmetric})-(\ref{eq:lambda_complete})
we see that if the motion of the perturber is both supersonic and 
super-Alfv\'enic, which corresponds to case IV,
the density disturbance is confined to a rear cone defined by
the condition $z<-|\gamma|R$.
In this regime, the surfaces of constant
density within the wake correspond to similar hyperbolae in the $z-R$ plane,
at the rear of the body,
with eccentricity $e={\mathcal{M}}/\sqrt{\xi}$.
From Eq.~(\ref{eq:xi_def}), it is simple to show that $\xi>1$ in this case.
Therefore, given a sonic Mach number ${\mathcal{M}}$,
the angular aperture of the cone is larger in the presence
of magnetic fields. 
In analogy to the unmagnetized medium, one could define the effective
speed of propagation of the disturbance as $v_{p}=\sqrt{\xi}c_{s}$ in case IV. 
At large Mach numbers (say ${\mathcal{M}}\gg \sqrt{\xi}$), 
the cone is very narrow ($e\gg 1$) and, as expected, 
the propagation speed coincides with the magnetoacoustic velocity, 
$v_{p}\simeq \sqrt{c_{s}^{2}+c_{A}^{2}}$. Hence, at these large 
${\mathcal{M}}$ values,
the stationary flow is similar to that in the nonmagnetic case but 
replacing the sound speed by the magnetosonic speed. 

Now consider case IV but when the perturber moves at the same velocity
as the effective speed of the disturbance, so that $V_{0}=v_{p}$ or, 
equivalently, ${\mathcal{M}}=\sqrt{\xi}$. In the non-magnetic case,
this condition corresponds to $\xi=1$ and, therefore, 
the velocity of the body is in
resonance with the sound speed in the medium. 
One could naively think that, at ${\mathcal{M}}=\sqrt{\xi}$,  
the response of the medium is maximum because of the resonance between
$V_{0}$ and $v_{p}$. 
This is not true for $\Upsilon>1$ because
${\mathcal{M}}=\sqrt{\xi}$ implies ${\mathcal{M}}=\Upsilon$
(using Eq.~\ref{eq:xi_def}), $\eta=1$ (from Eq.~\ref{eq:eta_def})
and, thereby, $\alpha=0$. We learn that a mass moving
at the Alfv\'en speed in a medium with $\Upsilon>1$, does not 
generate any density disturbance
in the ambient gas because the velocity field of the stationary flow 
is divergence-free ($\nab\cdot \vv'=0$).

\subsection{A comparison with Dokuchaev (1964)}

As already mentioned, Dokuchaev (1964) calculated, for the first time,
the properties of the wake created by a star moving along 
the field lines, by treating it as a linear perturbation.
His analysis started from the time-dependent linearized equations of
magnetohydrodynamics, including a source term $Q$ in
the continuity equation, representing the gas replenishment by the star. 
Although he used the time-dependent equations, he tacitly assumed that
the object's gravitational field is active since $t=-\infty$, so that
the wake is in a steady state.
For the case without mass injected by the star ($Q=0$), the
physical stand points used by Dokuchaev (1964) are exactly the
same as those adopted in \S \ref{sec:perturbed_density}, except that he chose
a reference frame in which the unperturbed background gas is at rest. 

Dokuchaev (1964) found closed fourth-order differential equations for $\rho$
and the radial component $v_{r}$ of the velocity. Through Fourier-Hankel
transformations, he was able to solve the differential equation for $\rho$.
He found a similar expression for $\rho$ as that given in 
Eq.~(\ref{eq:alpha_axisymmetric})
but he failed to separate correctly the different intervals 
for $\lambda$ (Eq.~\ref{eq:lambda_complete}) and the intervals at which the
isocontours are ellipsoids or hyperboloids. In particular, he claimed that
the isocontours are ellipsoids at any Mach number below 
$(c_{s}^{2}+c_{A}^{2})^{1/2}/c_{s}=(1+\Upsilon^{2})^{1/2}$,
which is misleading.

\subsection{Gravitational drag force in the axisymmetric case}
\label{sec:dragforce_axi}
Once we have the gas density enhancement $\alpha(\rr)$ in the ambient medium,
we can calculate the gravitational force exerted on the body by its
own wake:
\begin{equation}
\FF_{DF}=2\pi G M \rho_{0} \int \int dz \,dR \,\,R \,\alpha(\rr)
\frac{z} {(z^{2}+R^{2})^{3/2}} \hat{\zz}.
\end{equation}
The net drag is zero when the isodensity contours are ellipsoids, i.e.~when $\gamma^{2}>0$,
because the wake exhibits front-back symmetry. 
At values $\gamma^{2}<0$, however, the region
of perturbed density is confined to a cone  and the drag force is nonvanishing.
Evaluating the integrals in spherical coordinates ($R=r\sin \theta$ and
$z=r\cos\theta$), and using the variable $\mu$ defined
as $\mu=\cos\theta$, the drag force can be expressed as:
\begin{equation}
\FF_{DF}=(1-\eta)
\frac{4\pi G^{2} M^{2} \rho_{0}}{V_{0}^{2}} 
{\mathcal{I}}
\int_{r_{\rm min}}^{r_{\rm max}} \frac{dr}{r} \hat{\zz},
\end{equation}
where
\begin{equation}
{\mathcal{I}}=\frac{1}{2} \int_{-1}^{1} d\mu \frac{\mu\lambda{\mathcal{M}}^{2}/\xi}
{\sqrt{1-\xi^{-1}{\mathcal{M}}^{2}+\mu^{2}\xi^{-1}{\mathcal{M}}^{2}}}.
\end{equation}
As already said, the drag force is nonzero in cases II and IV, where 
$\gamma^{2}<0$ and thus $\xi>0$. 
In case II, $\lambda=2$ for all $\mu$ between 
$\mu_{\rm lower}=({\mathcal{M}}^{2}-\xi)^{1/2}/{\mathcal{M}}$ and $1$, so
that ${\mathcal{I}}=1$.
In case IV, $\lambda=2$ for all $\mu$
between $-1$ and $\mu_{\rm upper}=-({\mathcal{M}}^{2}-\xi)^{1/2}/{\mathcal{M}}$
and thus ${\mathcal{I}}=-1$.
Since $1-\eta <0$ in case II, we can write $(1-\eta){\mathcal{I}}=|1-\eta|$ 
and the resultant expression for the force is:
\begin{equation}
 \FF_{DF}= \left\{ \begin{array}{ll}
        - |1-\eta| {\mathcal{F}}\ln\Lambda \hat{\zz}  & \mbox{if ${\mathcal{M}}>{\rm max}(1,\Upsilon)$} \\
         & \mbox{or ${\mathcal{M}}_{\rm crit}<{\mathcal{M}}<{\rm min}(1,\Upsilon)$};\\
      0 & \mbox{otherwise} \end{array} \right.
\label{eq:axi_DF}
\end{equation}
where
\begin{equation}
{\mathcal{F}}=\frac{4\pi G^{2} M^{2} \rho_{0}}{V_{0}^{2}},
\label{eq:drag_dokuchaev}
\end{equation}
and $\ln \Lambda$ is the Coulomb logarithm.  
Note that the density diverges in the wake at Mach numbers close to
${\mathcal{M}}_{\rm crit}$ because $\xi\rightarrow 0$. However,
the drag force is finite 
because the opening angle of the cone becomes very narrow.
Still, the drag force peaks at 
Mach numbers close to ${\mathcal{M}}_{\rm crit}$ because the factor
$|1-\eta|/V_{0}^{2}$ in the formula for $\alpha$
increases when ${\mathcal{M}}$ decreases.

\begin{figure*}
  \plotone{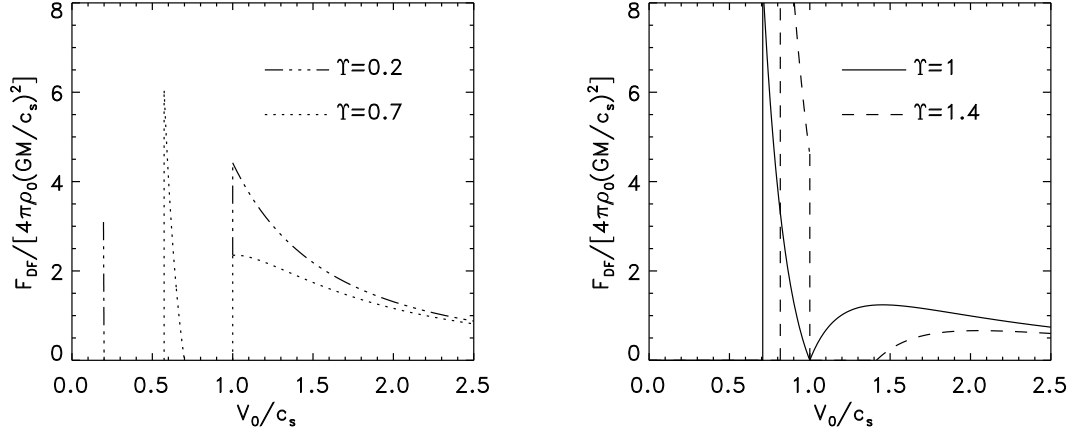}
  \caption{\label{fig:dokuchaev}
  Gravitational drag force as a function of Mach number, 
at $t=100 r_{\rm min}/c_{s}$,
as predicted by the steady-state linear-theory in the axisymmetric
case, for different values of $\Upsilon$.}
\end{figure*}

Figure \ref{fig:dokuchaev} shows the DF force felt by the body at 
$t=100 r_{\rm min}/c_{s}$, as a function of the Mach number
and for different values of $\Upsilon$.
In analogy to the unmagnetized case, we take $\Lambda=V_{0}t/r_{\rm min}=100 V_{0}/c_{s}$.
Dokuchaev (1964) claimed that the drag force
is nonzero only at ${\mathcal{M}}>(1+\Upsilon^{2})^{1/2}$
(see \S\ref{sec:magnetizedmedium}).
This is incorrect. For instance, for $\Upsilon=1$, 
the drag force is different from zero at ${\mathcal{M}}>{\mathcal{M}}_{\rm crit}=0.70$. 
If the ratio between the Alfv\'en and sound speeds is of $\Upsilon^{2}=2$,
the DF force is nonvanishing in the intervals $0.816<{\mathcal{M}}<1$, 
and ${\mathcal{M}}>1.41$.  In fact, there exists always
a subsonic velocity range at which the drag force is nonzero.

As long as $\Upsilon \neq 0$,
the DF force has two local maxima; one located at ${\mathcal{M}}_{\rm crit}$
and the other one at $\mach={\rm max}(1,\sqrt{2}\Upsilon)$.
The drag force strength at ${\mathcal{M}}_{\rm crit}$
increases with $\Upsilon$, whereas the drag force
at the second local maximum decreases with $\Upsilon$.
As Figure \ref{fig:dokuchaev} clearly shows, at low $\Upsilon$-values,
the width of the interval with $F_{DF}\neq 0$ around ${\mathcal{M}}_{\rm crit}$
becomes very narrow. 
For instance, the width of that interval is only of 
$4\times 10^{-3}$ for $\Upsilon=0.2$. 
Hence, the drag force at subsonic values is
irrelevant for astrophysical purposes when the Alfv\'en speed is
sufficiently small as compared to the sound speed.
For $\Upsilon>0.4$, the drag force at the local
maximum ${\mathcal{M}}_{\rm crit}$
is always larger than the drag force at the other local maximum
${\mathcal{M}}={\rm max}(1,\sqrt{2}\Upsilon)$. 
In the particular case of $\Upsilon=1.41$, the drag force 
is a factor $>6$ 
stronger in the interval $0.816<{\mathcal{M}}<1$ than it is at  ${\mathcal{M}}>1$.

\begin{figure}
\epsscale{1.2}
\plotone{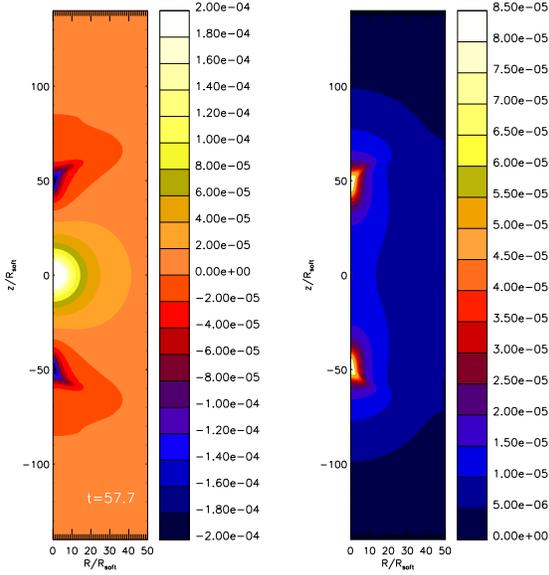}
\caption{\label{fig:map_0_141}
Color map of the density $\rho/\rho_{0}$ (left panel)
and magnetic field $B_{z}'/B_{0}$ (right panel), in the $(R,z)$-plane for
a case with $V_{0}=0$ and $\Upsilon=1.41$, in a natural logarithmic scale.}
\end{figure}

\begin{figure*}
  \epsscale{1.1}
  \plotone{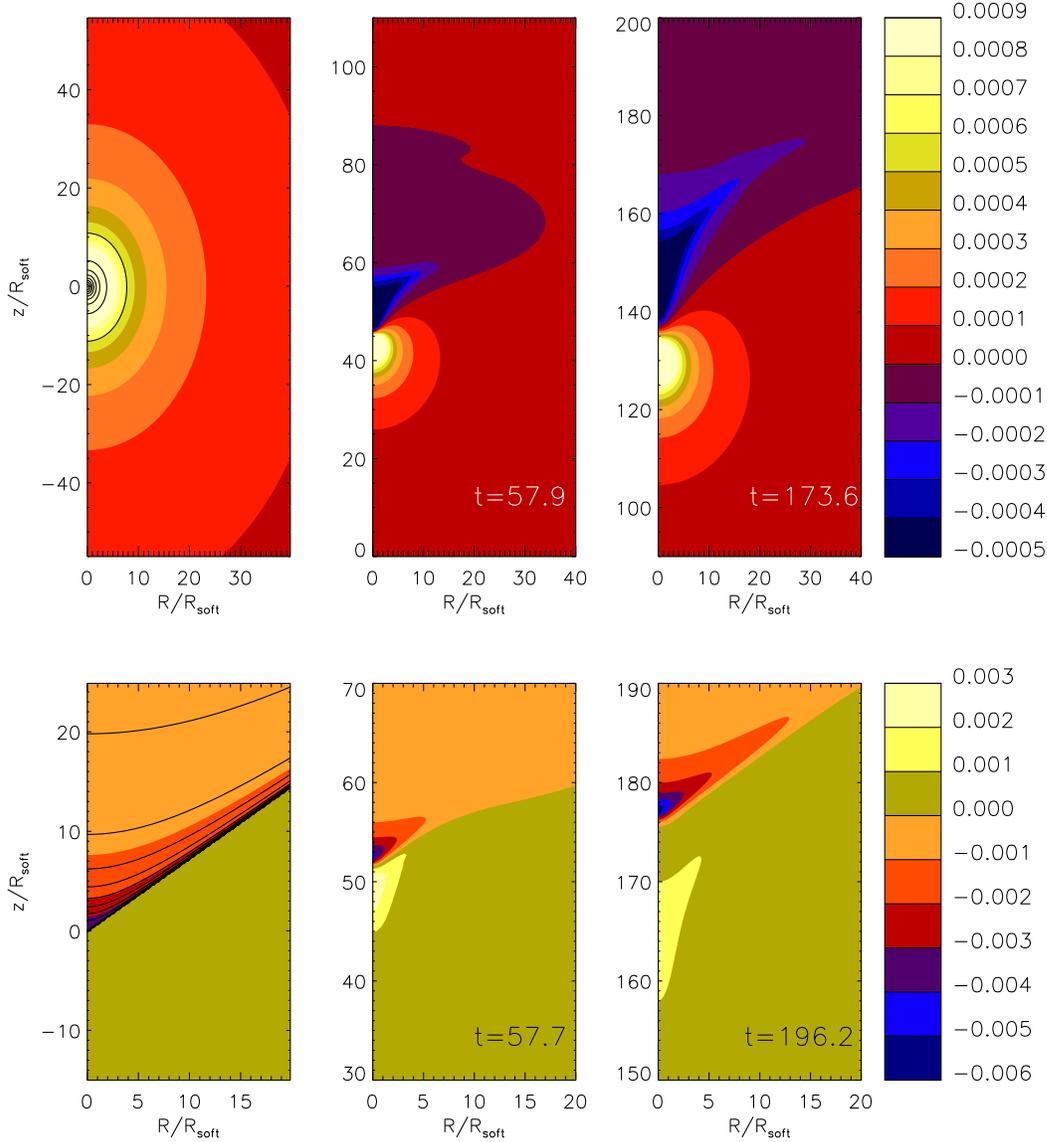}
  \caption{
Color map of the density $\rho/\rho_{0}$, in the $(R,z)$-plane,
 for the cylindrical case 
with ${\mathcal{M}}=0.75$ (upper panels)
and for ${\mathcal{M}}=0.9$ (lower panels), in a natural logarithmic
scale. Both models have $\Upsilon=1.41$, implying $\mach_{\rm crit}=0.816$. 
Therefore, $\mach=0.75$ falls into the interval I, while $\mach=0.9$ lies
in the interval II (see \S \ref{sec:perturbed_density}).
To easy comparison, the density map in the steady-state
for a perturber seated at $R=z=0$
is shown in the left panels. The central and right panels display
the density in the wake at two snapshots, 
when the perturber is dropped suddenly at $t=0$ at the origin of
the coordinate system. 
The time of the snapshots is given in the lower right hand corner of each
panel in units of $t_{\rm cross}$.
 }
\label{fig:map_07+09}
\end{figure*}

\begin{figure*}
  \epsscale{1.1}
  \plotone{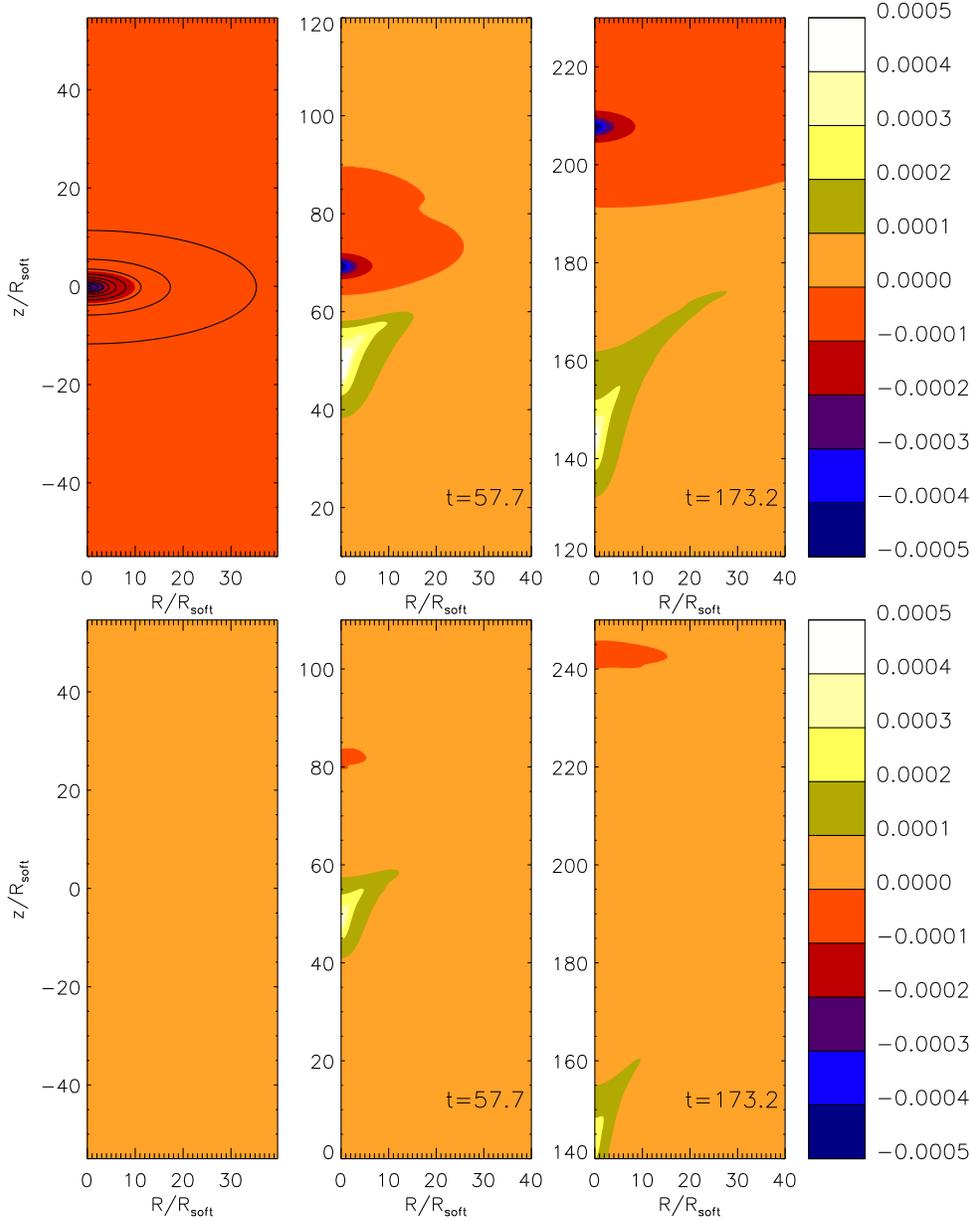}
  \caption{
Same as Fig.~\ref{fig:map_07+09} but for
${\mathcal{M}}=1.2$ (upper panels), which falls into the interval III,
and ${\mathcal{M}}=1.4$ (lower panels). Again $\Upsilon=1.41$.}
\label{fig:map_12+14}
\end{figure*}

\begin{figure*}
  \epsscale{0.87}
  \plotone{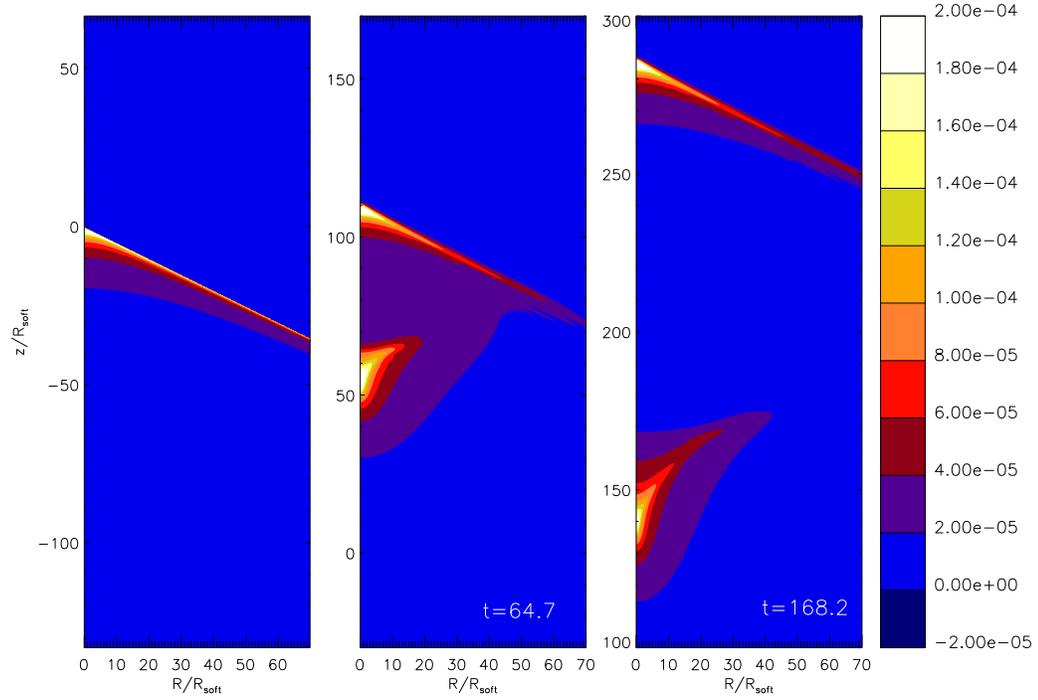}
  \caption{
Same as Fig.~\ref{fig:map_07+09} but
for ${\mathcal{M}}=1.7$. Again $\Upsilon=1.41$. This Mach number
lies in the interval IV.}
\label{map_17}
\end{figure*}

\begin{figure*}
\epsscale{0.9}
\plotone{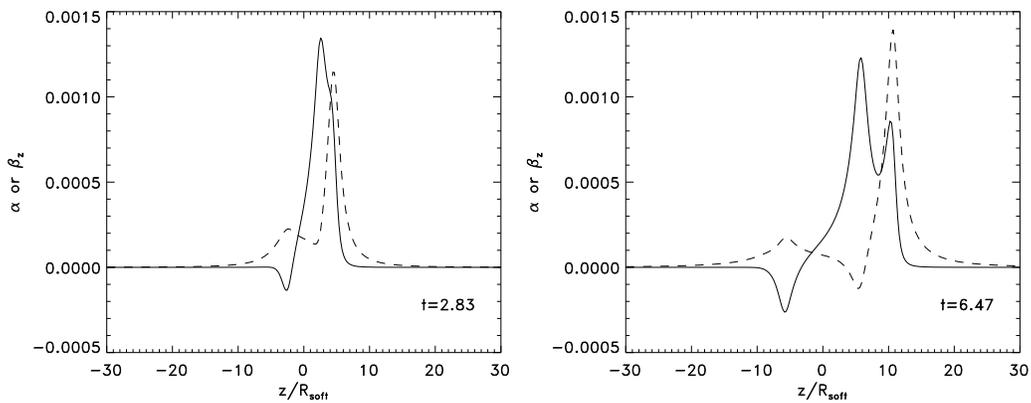}
\caption{Distributions of the perturbed density $\alpha$
(solid lines) and the $z$-component of the perturbed magnetic field 
(dashed lines) along a cut
at $R=0$ at two different times, for the same model as that shown in
Fig.~\ref{map_17} ($\mach=1.7$ and $\Upsilon=1.41$). }
\label{fig:tulip}
\end{figure*}

\begin{figure*}
  \epsscale{0.87}
  \plotone{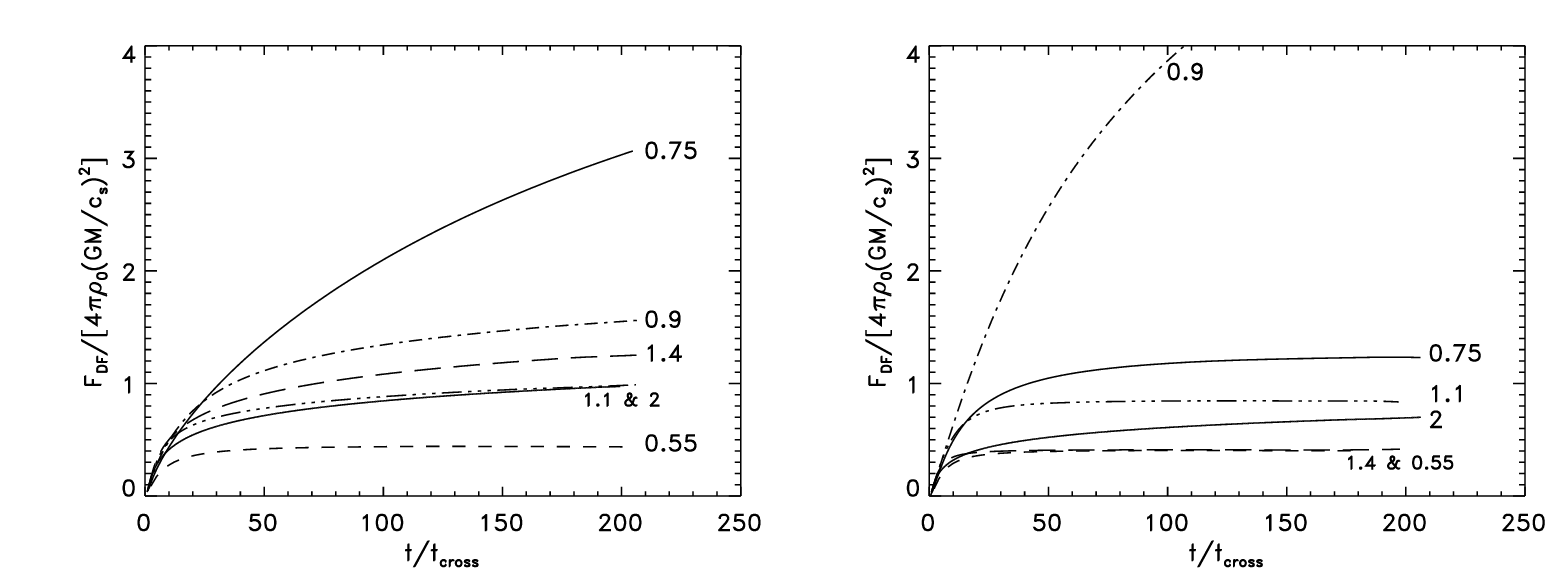}
  \caption{\label{fig:force_time}
  Temporal evolution of the gravitational drag force in the
axisymmetric model, at six different
Mach numbers between $0.55$ and $2$, 
for $\Upsilon=1$ (left panel), and for $\Upsilon=1.41$ (right panel).
The number on each curve is the Mach number $\mach$.}
\end{figure*}

\section{Time-dependent equations}
\label{sec:time_dependent}
The steady state analysis in the axisymmetric
case predicts zero drag force at certain Mach numbers 
because the perturber is surrounded by complete ellipsoids that 
exert no net force. 
As Ostriker (1999) demonstrated in the field-free case, the time-dependent
analysis in which the body is dropped suddently at $t=0$ allows 
to capture the asymmetric density shells in the far 
field which exert a gravitational drag on the body.
Other advantage of the time-dependent approach is that, contrary
to what happens when assuming steady-state, the
ambiguity in the definition of the maximum cut-off distance
$r_{\rm max}$ is fixed.

Without loss of generality, it is convenient to use the gas frame of reference
in which the ambient gas is initially at
rest, the initial magnetic field is along the $z$-axis and the body 
moves with velocity $V_{0,y}\hat{\yy}+V_{0,z}\hat{\zz}$.
The first order continuity equation is 
\begin{equation}
\frac{\partial \rho'}{\partial t}+\rho_{0} \nab\cdot \vv'=0,
\label{eq:continuity}
\end{equation}
the MHD Euler equation
\begin{equation}
\frac{\partial \vv'}{\partial t}=
-\frac{c_{s}^{2}\nab\rho'}{\rho_{0}}
-\nab\Phi + \frac{1}{4\pi\rho_{0}} (\nab\times \BB')\times
\BB_{0},
\label{eq:MHDEuler}
\end{equation}
and the induction equation:
\begin{equation}
\frac{\partial \BB'}{\partial t}=\nab \times (\vv'\times \BB_{0}).
\label{eq:Induction}
\end{equation}
The medium initially uniform will respond to the gravitational pull of the
body through the emission of fast and slow Alfv\'en waves and sound waves.
In the following we will manipulate the above equations to obtain a closed
system of two differential equations for $\rho'$ and $B_{z}'$ in analogy to  
the steady-state case.

Using Eq. (\ref{eq:div_Lorentz}) in the divergence of Equation (\ref{eq:MHDEuler})
\begin{equation}
\frac{\partial (\nab\cdot \vv')}{\partial t}
=-\frac{c_{s}^{2}}{\rho_{0}}\nabla^2 \rho' -\nabla^{2} \Phi-
\frac{B_{0}}{4\pi \rho_{0}}\nabla^{2}B_{z}'.
\label{eq:divergence}
\end{equation}
By substituting Eq.~(\ref{eq:continuity}) into Eq.~(\ref{eq:divergence}), 
we obtain
\begin{equation}
\frac{1}{\rho_{0}} \frac{\partial^{2}\rho'}{\partial t^{2}}
=\frac{c_{s}^{2}}{\rho_{0}}\nabla^2 \rho' +\nabla^{2} \Phi+
\frac{B_{0}}{4\pi \rho_{0}}\nabla^{2}B_{z}'.
\end{equation}
In terms of $\alpha$ and $\beta_{z}$, it yields
\begin{equation}
\frac{\partial^{2}\alpha}{\partial t^{2}}
=c_{s}^{2}\nabla^2 \alpha +\nabla^{2} \Phi+
c_{A}^{2}\nabla^{2}\beta_{z}.
\label{eq:weq1}
\end{equation}
Here, the magnetic effect on the density perturbation appears as a
inhomogeneous term. We may recover the classical non-magnetic equation 
for $\alpha$ by taking $c_{A}=0$. 

On the other hand, the third component of the induction equation (Eq.~\ref{eq:Induction}) implies:
\begin{equation}
\frac{\partial B_{z}'}{\partial t}=-B_{0} \left(\frac{\partial v_{x}'}{\partial x}
+\frac{\partial v_{y}'}{\partial y}\right).
\label{eq:bzt}
\end{equation}
Equations (\ref{eq:continuity}) and (\ref{eq:bzt}) give
\begin{equation}
\frac{\partial B_{z}'}{\partial t}=B_{0}\left(\frac{1}{\rho_{0}}\frac{\partial
\rho'}{\partial t}+\frac{\partial v_{z}'}{\partial z}\right).
\label{eq:bz_t}
\end{equation}
From the third component of the equation of motion (Eq.~\ref{eq:MHDEuler})
\begin{equation}
\frac{\partial v_{z}'}{\partial t}=-\frac{c_{s}^{2}}{\rho_{0}}\frac{\partial 
\rho'}{\partial z}-\frac{\partial \Phi}{\partial z},
\end{equation}
we know that 
\begin{equation}
\frac{\partial}{\partial t}\left(
\frac{\partial v_{z}'}{\partial z}\right)=
-\frac{c_{s}^{2}}{\rho_{0}}\frac{\partial^{2}
\rho'}{\partial z^{2}}-\frac{\partial^{2} \Phi}{\partial z^{2}}.
\label{eq:vz_tz}
\end{equation}
Inserting Eq.~(\ref{eq:vz_tz}) into the temporal derivative of 
Eq.~(\ref{eq:bz_t}), one finds
\begin{equation}
\frac{1}{B_{0}}
\frac{\partial^{2} B_{z}'}{\partial t^{2}}= 
\frac{1}{\rho_{0}}\frac{\partial^{2}\rho'}
{\partial t^{2}}-\frac{c_{s}^{2}}{\rho_{0}} \frac{\partial^{2}\rho'}
{\partial z^{2}}-\frac{\partial^{2}\Phi}{\partial z^{2}}.
\end{equation}
In dimensionless form:
\begin{equation}
\frac{\partial^{2}\beta_{z}}{\partial t^{2}}=
\frac{\partial^{2} \alpha}{\partial t^{2}}-c_{s}^{2}
\frac{\partial^{2} \alpha}{\partial z^{2}}-
\frac{\partial^{2}\Phi}{\partial z^{2}}.
\label{eq:weq2}
\end{equation}
Putting together, the equations (\ref{eq:weq1}) and (\ref{eq:weq2}) to solve can
be written as 
\begin{equation}
\Box_{s}\alpha = \nabla^{2}\tilde{\Phi}+
\Upsilon^{2}\nabla^{2}\beta_{z},
\label{eq:wave_equation1}
\end{equation}
\begin{equation}
\Box_{A}\beta_{z} = \Upsilon^{-2}
\left(\frac{\partial^{2}}{\partial x^{2}}+\frac{\partial^{2}}{\partial y^{2}}
\right) (\alpha+\tilde{\Phi}),
\label{eq:wave_equation2}
\end{equation}
where $\tilde{\Phi}$ is the gravitational potential in units of $c_{s}^{2}$
(i.e.~$\tilde{\Phi}=\Phi/c_{s}^{2}$) and
we have used
the Lorentz invariant D'Alembertian $\Box$, defined as:
\begin{equation}
\Box_{l} \phi = \left(\frac{1}{c_{l}^{2}}\frac{\partial^{2}}{\partial t^{2}}
-\nabla^{2}\right)\phi.
\end{equation}
The first equation (Eq.~\ref{eq:wave_equation1}) governs the evolution
of the density in the presence of a gravitational potential and magnetic 
fields. The second equation (Eq~\ref{eq:wave_equation2}) describes
the evolution of a frozen-in magnetic field when the gas is subject
to pressure gradients and to an external gravitational potential.
The inhomogeneous term in Eq.~(\ref{eq:wave_equation2}) does not
have $z$-derivatives because 
gas motions in that direction does not
compress, stir or stretch the background magnetic field.
The resulting equations (\ref{eq:wave_equation1}) and (\ref{eq:wave_equation2}) 
conform to a set of two coupled non-homogeneous wave equations.
For a point-mass perturber, it is simple to find $\alpha$ in 
the Fourier-Laplace space, 
$\hat{\alpha}(\kk,\omega)$, but the inverse Fourier-Laplace integral
cannot be given in a closed analytic form.

In order to gain physical insight, consider first a two-dimensional example. 
If $\Phi=\Phi(x,y)$, that is, if the perturber is an infinite cylinder 
with a certain radial density profile $\rho_{p}=\rho_{p}(R)$,
then $\beta_{z}=\alpha$ because of the flux-freezing condition, 
and $\alpha$ satisfies a simple wave equation
with magnetoacoustic speed:
\begin{equation}
\left(\frac{1}{c_{s}^{2}+c_{A}^{2}}\frac{\partial^{2}}{\partial t^{2}}
-\nabla^{2}\right)\alpha=\frac{1}{1+\Upsilon^{2}}\nabla^{2}\tilde{\Phi}.
\end{equation}
The physical reason is that motions are always perpendicular to the frozen-in
magnetic field lines. Magnetohydrodynamical equilibrium is reached 
within the magnetosonic cylinder. At a later stage, Parker
instabilities can develop (S\'anchez-Salcedo \& Santill\'an 2011).

In the purely hydrodynamical problem, the equation governing the evolution
of $\alpha$ is $\Box_{s}\alpha=\nabla^{2}\tilde \Phi$. If the perturber
is a point source, we have 
$\Box_{s}\alpha=(4\pi G M/c_{s}^{2})\delta(x-V_{0,x}t)
\delta(y-V_{0,y}t)\delta(z-V_{0,z}t)H(t)$. Hence, the density remains
unperturbed outside the causal region for sound waves (see Ostriker 1999).
In a magnetized medium, however, the situation is different because
Equation (\ref{eq:wave_equation2}) for
the perturbed magnetic field $\beta_{z}$ has a source term
($\partial^{2}\Phi/\partial x^{2}+\partial^{2}\Phi/\partial y^{2}$)
which does not vanish even outside the causal region for 
magnetosonic waves.

In the next Section we will solve the coupled wave-equations numerically.
To do so, the perturber gravitational potential will be
modeled by a smooth core Plummer potential:
\begin{equation}
\Phi(\rr,t)=-\frac{GM{\mathcal{H}}(t)}{\sqrt{x^{2}+(y-V_{0,y}t)^{2}+(z-V_{0,z}t)^{2}+
R_{\rm soft}^{2}}},
\end{equation}
where $R_{\rm soft}$ is the softening radius and ${\mathcal{H}}$ is
a Heaviside step function. 
Stellar and globular clusters can be accurately described by Plummer potentials.
These type of models were also 
used in S\'anchez-Salcedo \& Brandenburg (1999, 2001), Kim \& Kim (2009),
and Kim (2010) to study DF.

\begin{figure*}
  \plotone{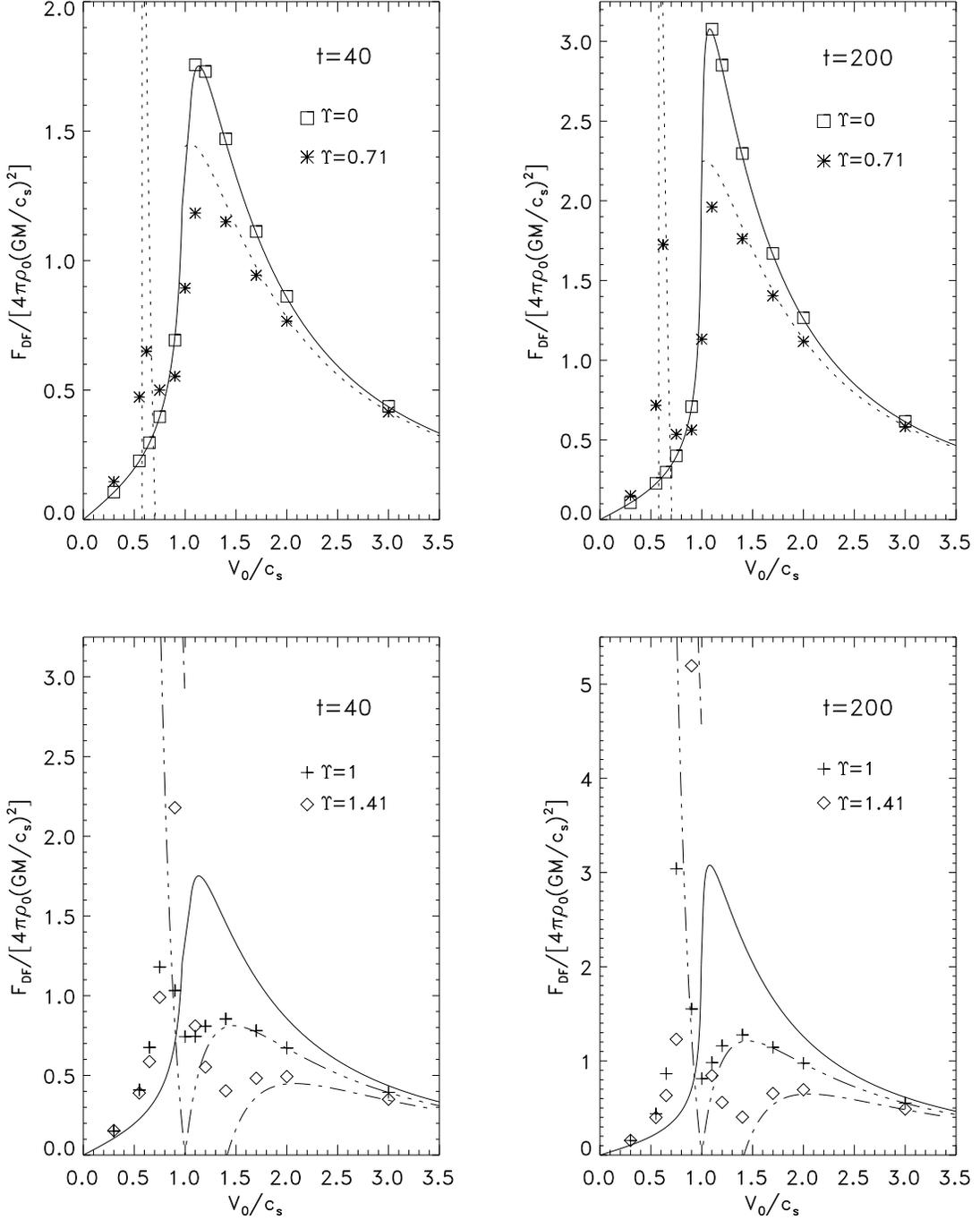}
  \caption{\label{fig:fig_200_0g}
  Gravitational drag force for the time-dependent axisymmetric models 
against the Mach number at $t=40t_{\rm cross}$ (left panel)
and at $t=200t_{\rm cross}$ (right panel). 
Symbols correspond to the numerical models.
The solid curves plot Ostriker's formula, while the remainder curves
draw the drag force as predicted by Equation (\ref{eq:axi_DF}),
adopting $r_{\rm min}=2.25R_{\rm soft}$ and $r_{\rm max}=V_{0}t$.
}
\end{figure*}

\begin{figure*}
  \epsscale{0.87}
  \plotone{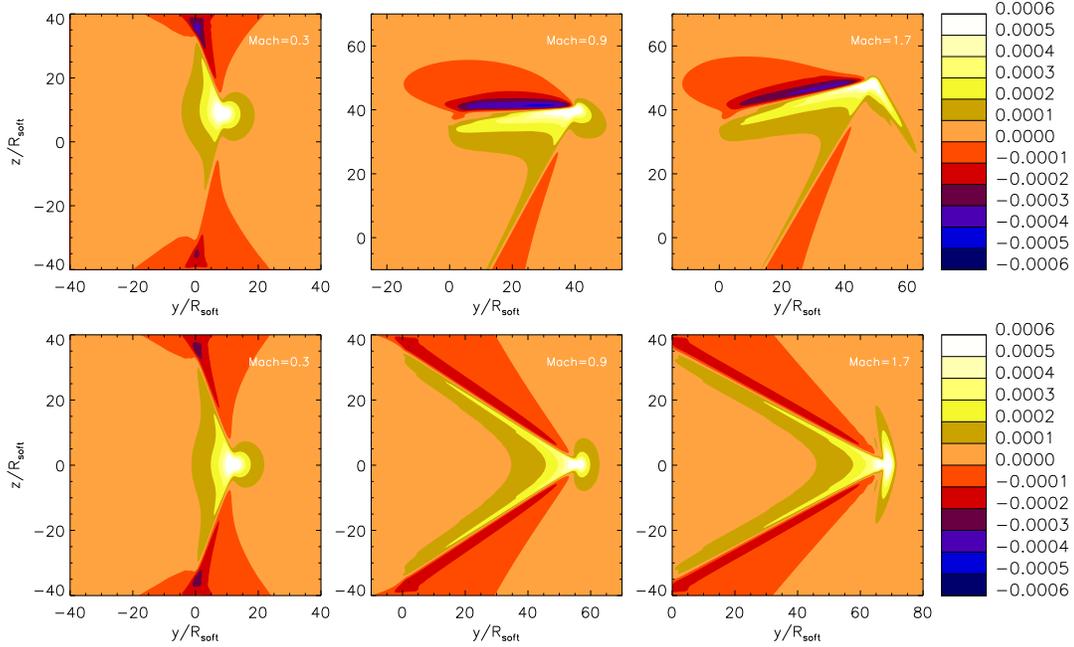}
  \caption{\label{fig:fig_map_45+90}
Snapshots of the 
density map along a cut-off through the $(y,z)$-plane at $x=0$, in natural logarithmic scale,
for $\Theta=45^{\circ}$ (upper panels) and for $\Theta=90^{\circ}$
(lower panels), at three different Mach numbers.
The Mach number $\mach$ is indicated at the right corner in each panel.
The perturber was dropped at $t=0$ and moves on a rectilinear orbit in the $(y,z)$-plane.
 In all panels, $\Upsilon=1.41$.
}
\end{figure*}

\begin{figure}
  \epsscale{0.87}
  \plotone{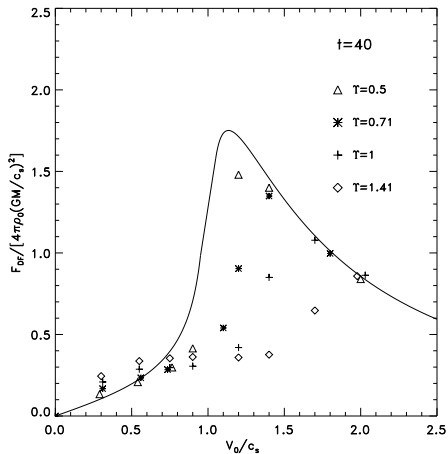}
  \caption{\label{fig:fig_90g}
  Gravitational DF force as a function of Mach number for different $\Upsilon$-values
  at $t=40t_{\rm cross}$. The perturber moves perpendicular to the magnetic field lines 
  (i.e.~$\Theta=90^{\circ}$).
  The solid curve plots Ostriker's formula, which was derived for unmagnetized media,
   with $r_{\rm min}=2.25R_{\rm soft}$ and $r_{\rm max}=V_{0}t$. To make the plot
   readable, the symbols at
   $\mach=0.3, 0.55, 0.75$ and $2$ have been slightly shifted in the horizontal direction.
}
\end{figure}

\begin{figure}
  \epsscale{0.77}
  \plotone{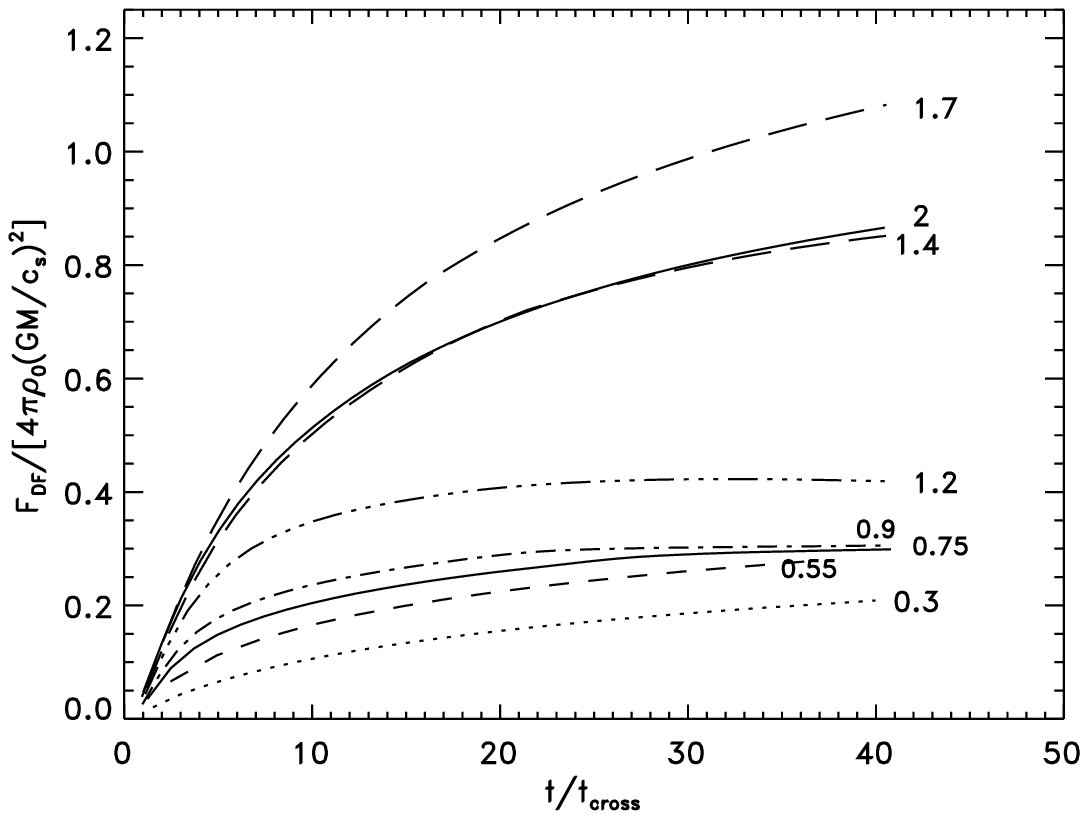}
  \caption{\label{fig:force_time_90g}
  Temporal evolution of the gravitational drag force for $\Upsilon=1$. 
  The perturber moves in a direction perpendicular to the magnetic field lines.
The numbers given at each curve represents the Mach number. 
}
\end{figure}

\section{Results}
\label{sec:results}
The coupled inhomogeneous wave equations (\ref{eq:wave_equation1}) and 
(\ref{eq:wave_equation2}) were solved using a 
finite difference scheme in a uniform grid.
The scheme is second order in space and third order in time. 
The temporal algorithm was
described in S\'anchez-Salcedo \& Brandenburg (2001). 
Calculations start with a uniform background density and magnetic field,
and the body is initially placed at the origin of the coordinate system with 
velocity $V_{0,y}\hat{\yy}+V_{0,z}\hat{\zz}$.
For the axisymmetric case, which occurs when $V_{0,y}=0$, the
calculations were carried out on a two-dimensional $(R,z)$-plane
in cylindrical symmetry.
In the general three-dimensional case ($V_{0,y}\neq 0$),
the variables $\alpha$ and $\beta_{z}$ are symmetric about the plane
$x=0$. Hence, we considered a finite domain with
$x\in [0, L_{x,{\rm max}}]$ and used symmetric boundary conditions at $x=0$, and
outflow boundary conditions in the other five caps of the computational
domain. However, the size of the domain was
taken large enough to ensure that the perturbed density and magnetic field
do not reach the boundaries.

As a test of the algorithm, we
studied the convergence of homogeneous wave modes by perturbing a uniform background medium.
We further tested convergence of our models for several resolutions
and found that four zones per $R_{\rm soft}$ suffice to have converged results.

We take $R_{\rm soft}$, $c_{s}$, and $t_{\rm cross}=R_{\rm soft}/c_{s}$ as the units of
length, velocity, and time, respectively.
A model can thus be specified with four dimensionless parameters:
$GM/(c_{s}^{2}R_{\rm soft})$, $\mach$, $\Upsilon$ and
$\Theta\equiv {\rm atan}(V_{0,y}/V_{0,z})$. $\Theta$ is the angle between $\VV_{0}$
and $\BB_{0}$.
Fixed ${\mathcal{M}}$, $\Theta$ and $\Upsilon$, the variables
$\alpha$ and $\beta_{z}$ depend
linearly on $GM/(c_{s}^{2}R_{\rm soft})$. Hence, in our calculations
we always take $GM/(c_{s}^{2}R_{\rm soft})=0.01/3$ and explore
how the density and the magnetic field in the wake depend on the other
three parameters.

\subsection{Axisymmetric case}
\label{sec:axisymmetriccase}

We first run models with the magnetic field terms swich off and compare the density
enhancement and the gravitational drag with previous linear calculations 
in Ostriker (1999) and S\'anchez-Salcedo \& Brandenburg (1999). 
We found excellent agreement, backing up our numerical model.
In the following, we will present results for a body moving along the field
lines of the unperturbed magnetic field, which corresponds to $\Theta=0$.

The simplest scenario occurs when the gravitational perturber is dropped
at $t=0$ at rest ($V_{0}=0$). As discussed at the
begining of \S \ref{sec:physical_interpretation}, 
the steady-state density distribution is identical
as that without any magnetic field. However, the inital relaxation stage
and the far density distribution are sensitive to magnetic effects. In fact,
while the problem has spherical symmetry at any time in the purely
hydrodynamical case,
this symmetry is broken in a magnetized medium because the
magnetic field dictates a preferential direction. 
Figure \ref{fig:map_0_141} shows maps of density and $B_{z}'$
for a case with $\Upsilon=1.41$. We see that the density
distribution in the vicinity of the body is indeed spherically symmetric
 and the magnetic field
takes essentially its initial value, implying that this part has reached
hydrostatic equilibrium. However, there are two symmetric underdense regions
along the $z$-axis in the outer parts. Physically, the origin of them
is that the magnetic field reduces the flow convergence toward the
symmetry axis 
because magnetic forces mainly affect the radial component of 
the velocity $v_{R}$.
This loss of radial convergence produces a wave with  
negative density enhancement ($\alpha<0$)
but a positive magnetic enhancement ($\beta_{z}>0$) 
because of the compression of
the magnetic field lines in the radial direction. 

Figure \ref{fig:map_07+09} shows snapshots of the density at the  
$(R,z)$ plane for $\Upsilon=1.41$ and
two subsonic velocities (${\mathcal{M}}=0.75$ and ${\mathcal{M}}=0.9$). 
In the time-dependent analysis, both cases present a region of negative density
enhancement (i.e.~$\alpha<0$) at the head of the perturber. 
As $\mach$ increases from $0$ to $0.9$, the underdense region at the 
head of the body
remains and gets deeper, while the underdensity in the downstream
region 
becomes less pronounced. This is simply consequence of the Doppler
effect; gradients become steeper
upstream (remind that this also occurs in the purely hydrodynamic case).
By comparing the density at two times (the central
and the right panels of Figure \ref{fig:map_07+09}), 
we see that the evolution of the density looks self-similar.

At ${\mathcal{M}}=0.75$, the steady-state analysis predicts a null 
drag force because of the front-back symmetry in the wake 
(see upper left panel of Fig.~\ref{fig:map_07+09}). 
In the finite-time case,
complete ellipsoids are visible only in the vicinity of the body. 
For instance, at $t=173.6t_{\rm cross}$ 
(with $\Upsilon=1.41$ and $\mach=0.75$), the
isodensity contours are not longer ellipsoids at distances beyond 
$\simeq 8 R_{\rm soft}$ from the body's center.
For comparison, in the absence of magnetic fields ($\Upsilon=0$) 
and at $\mach=0.75$,
ellipsoids within a radius of $43R_{\rm soft}$ are complete at 
$t=173.6t_{\rm cross}$. 
This difference is a consequence of the coupling between $\alpha$
and $\beta_{z}$.
At $z=136R_{\rm soft}$ and $R=0$ (i.e.~in the symmetry axis),
a steep front separates the fluid into two regions; one with $\alpha>0$ and
$\beta_{z}<0$, from another with $\alpha<0$ and $\beta_{z}>0$ 
This front leads to a deceleration of the gas, which expands radially.

At ${\mathcal{M}}=0.9$, a cone of negative density 
enhancement is located at the head 
of the perturber (see Fig.~\ref{fig:map_07+09}), as that predicted in \S\ref{sec:perturbed_density}, 
and a region of positive density enhancement at the rear. 
Because of the back-reaction of the magnetic field, the isodensity
contours of the tail are not incomplete ellipsoids at all. 
Note that the body is at the apex of the cone,
whereas the overdensity is detached from the perturber. It is important to remark that,
according to the analysis in \S\ref{sec:dragforce_axi} and Figure \ref{fig:dokuchaev},
the maximum drag force for $\Upsilon=1.41$ occurs at a Mach number 
close to $0.9$.

Figure \ref{fig:map_12+14} displays density maps for $\Upsilon=1.41$ and two supersonic 
Mach numbers: ${\mathcal{M}}=1.2$ (sub-Alfv\'enic perturber)
and ${\mathcal{M}}=1.4$ (trans-Alfv\'enic perturber). 
In the first case, the steady-state analysis predicts ellipsoidal
isocontours. In the time-dependent case, however,
ellipsoids are incomplete far enough away upstream from the perturber 
and, again, an overdensity wave at the rear of the perturber
moves away from the body. 
When the pertuber moves at $\mach=1.4$, a teneous ellipsoidal 
{\it underdense} envelop is still visible.

The overdensity behind the body has the 
shape of a tulip.
This tulip-shaped overdensity also appears at the rear of the body 
at ${\mathcal{M}}=1.4$ and at $\mach=1.7$ (see Figs.~\ref{fig:map_12+14}
and \ref{map_17}). 
A feature of the tulip-shaped overdensity wave is that it has a negative
magnetic enhancement. The tulip-shaped overdensity is a consequence
of our initial conditions and, as expected, it is detached from the body.
In order to illustrate the birth of the tulip-shaped wave, 
Figure \ref{fig:tulip}
shows the density and magnetic perturbations along the symmetry axis
for $\mach=1.7$,
at two early times. Initially,
$\beta_{z}$ increases due to the compression of magnetic field lines
(see the panel at $t=2.83t_{\rm cross}$). At the far edge of the tail,
$z\simeq -3R_{\rm soft}$, an underdense region with positive $\beta_{z}$ 
appears.  Later on (see the profiles at $t=6.47t_{\rm cross}$), the 
overdensity loses gravitational 
support and expands behind the body, decreasing the magnetic field strength,
until the magnetic pressure plus the magnetic tension provides 
sufficient radial confinement to the tulip-shaped structure, allowing
it to remain over long times.

In Figure \ref{map_17} it is simple to identify the 
modified Mach cone dragged by its point by the perturber. 
At $t=64.7t_{\rm cross}$, 
the Mach cone at the rear of the body is well-defined.
 Clearly, the timescale for the development of the Mach cone at the 
rear for ${\mathcal{M}}=1.7$
 is shorter than the timescale to form the upstream Mach front at ${\mathcal{M}}=0.9$. In fact, 
 Figure \ref{fig:map_07+09} shows that the
 cone for ${\mathcal{M}}=0.9$ is not well developed at $t=57.7t_{\rm cross}$.

Figure \ref{fig:force_time} shows the gravitational DF 
drag as a function of time for 
${\mathcal{M}}=0.55, 0.75, 0.9, 1.1, 1.4$ and $2$.
For $\Upsilon=1$ and ${\mathcal{M}}=0.55$, the drag force clearly saturates in 
$\sim 30t_{\rm cross}$.
In the remainder cases the drag force increases
with time. However, the drag force on a perturber moving in a medium with $\Upsilon=1.41$,
saturates for ${\mathcal{M}}=0.55, 0.75, 1.1$, and $1.4$.  
This means that the DF force saturates to a constant value at those
Mach numbers that the steady-state analysis predicts a null drag force.

\begin{figure*}
  \epsscale{0.87}
  \plotone{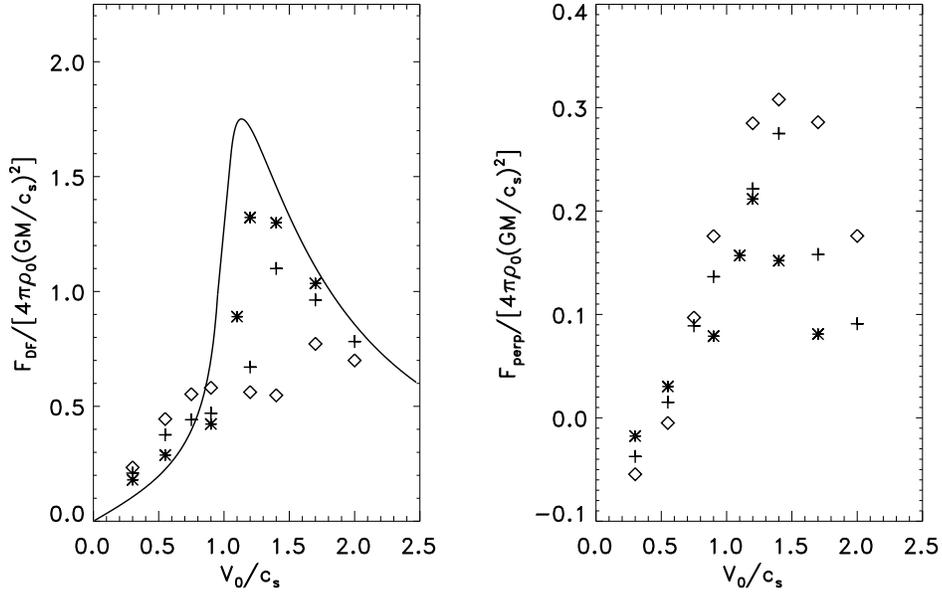}
  \caption{\label{fig:fig_45g}
 Components of the gravitational DF force 
parallel (left panel) and perpendicular (right panel) 
 to the direction $\VV_{0}$ (at $t=40t_{\rm cross}$) versus $\mach$, for 
 three different values of $\Upsilon$ ($\Upsilon=0.71$, asterisks; $\Upsilon=1$, crosses;
 $\Upsilon=1.41$, diamonds). 
  The angle between the perturber velocity and the magnetic field
is $45^{\circ}$. The key to symbols is the same as in Figure \ref{fig:fig_90g}.
}
\end{figure*}

In Fig.~\ref{fig:fig_200_0g} we plot the drag force at $t=40 t_{\rm cross}$ and 
$t=200 t_{\rm cross}$ for different values
of $\Upsilon$, together with the predicted force with $r_{\rm min}=2.25
R_{\rm soft}$. We see that for $\Upsilon=0$, there is a perfect agreement
between the Ostriker formula and the inferred values, confirming
the result that $r_{\rm min}=2.25R_{\rm soft}$ reported in 
S\'anchez-Salcedo \& Brandenburg (1999). 
The drag force formula given in Eq.~(\ref{eq:drag_dokuchaev}), with
$r_{\rm min}=2.25 R_{\rm soft}$, overestimates the drag
force in a neighbourhood of ${\mathcal{M}}_{\rm crit}$, where the 
drag force as a function of ${\mathcal{M}}$, becomes very cuspy.
As expected, the steady-state formula is more accurate at long timescales, except when
it predicts a zero net force. Roughly speaking, we may say that, for the axisymmetric case,
the gravitational drag in a magnetized medium is always smaller or equal
as the drag force in the unmagnetized case for supersonic perturbers 
($\mach>1$),
whereas the drag is always larger or equal in a magnetized medium
at subsonic perturbers ($\mach<1$).

\begin{figure}
  \epsscale{0.87}
  \plotone{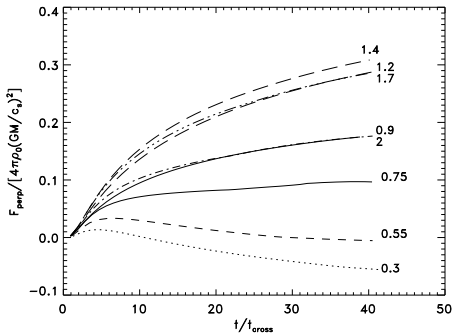}
  \caption{\label{fig:fig_force_perp_45g}
  Component of the DF force perpendicular to the direction of motion
as a function of time for $\Upsilon=1.41$ and $\Theta=45^{\circ}$. 
The number on each curve is the Mach number. 
}
\end{figure}

\subsection{Magnetic field perpendicular to the direction of motion of the perturber}
In \S \ref{sec:axisymmetriccase}, we have focused on the case where
the angle formed between the velocity of the perturber and the magnetic field,
$\Theta$, was equal to $0$. In such a situation, the problem has axial symmetry and it
is possible to find the analytical solution in the steady state.
However, it is by no means clear how the gravitational drag force depends on $\Theta$.
A visual comparison of the density wake structures 
for $\Theta=0$, $\Theta=\pi/4$ and $\Theta=\pi/2$ in Figures 
\ref{fig:map_07+09}
and \ref{fig:fig_map_45+90} would lead us to think that the resulting wake
at $\Theta=\pi/4$ is more similar to the case with $\Theta=\pi/2$ than to $\Theta=0$.
In particular, we would like to stress the remarkably different structure of the wake
for $\Theta=0$ (Fig.~\ref{fig:map_07+09}) and $\Theta=\pi/2$ (Fig.~\ref{fig:fig_map_45+90})
at $\mach=0.9$.

In this section we will discuss in detail the extreme case where the perturber
moves perpendicular to the field lines. In Appendix \ref{sec:appendix2}, 
the perturbed density is given in Fourier space. 
At subsonic Mach numbers and $\Theta=\pi/2$, the perturber is surrounded by
a ellipsoid-like envelope but also presents a tail with positive and 
negative $\alpha$-values
separated by a sharp front (see Fig.~\ref{fig:fig_map_45+90} for 
$\Upsilon=1.41$). 
Approaching to the upstream axis of motion of the body, the $y$-axis,
the plowing up of field lines increases the total pressure. At
low Mach numbers, say $\mach=0.3$, underdense regions are now formed  
in the direction of the ambient field lines, which are lagged behind 
the body (remind that regions
with negative $\alpha$ appear along the field lines; 
see Fig.~\ref{fig:map_0_141}).
At $\mach=0.9$, even if the motion is subsonic and sub-Alfv\'enic, 
a magnetic bow wave with sharp edges and opening angle 
$\arctan [c_{A}/V_{0}]$ is apparent in 
Fig.~\ref{fig:fig_map_45+90}. Note that when we say that it is 
sub-Alfv\'enic, we only mean $c_{A}/V_{0}<1$. However, some caution
should be used when interpreting this ratio because the velocities
are oriented in different directions. Since the velocity of the
perturber is always orthogonal to the ambient direction of propagation
of Alfv\'en waves, the Alv\'en speed in the direction of motion
of the perturber is zero and, thus, the body is always infinitely 
super-Alfv\'enic in the direction of motion. 
The morphology of the wake is the result of a competition between 
the gravitational
focusing of gas by the perturber and the drainage of gas along
magnetic field lines. 
We should warn here that the wake is not axisymmetric and thus the density
map in the $(x,y)$-plane is different than the map in the $(y,z)$-plane.

When the perturber travels faster than the magnetosonic velocity
$c_{s}(1+\Upsilon^{2})^{1/2}$, a magnetosonic Mach cone is 
formed at the rear of the pertuber; the entire perturbed density 
distribution lags the perturber.
In the $(y,z)$-plane, the perturber creates two magnetic bow waves;
the Alfv\'enic wave with opening angle $\arctan [c_{A}/V_{0}]$ and 
the magnetosonic wave with opening
angle $\arctan [(c_{A}^{2}+c_{s}^{2})^{1/2}/V_{0}]$.   

The gravitational drag force is the result of the contribution of all the
parcels in the domain and it is not possible to estimate its value just by comparing the density
structure by eye. 
Figure \ref{fig:fig_90g} shows the gravitational drag as a function of perturber's Mach number
for different values of $\Upsilon$, together with the gravitational drag 
in the unmagnetized case using Ostriker's formula. All the points at Mach
number larger than $0.7$ lie on or below Ostriker's curve,
implying that at $\mach>0.7$ the drag force is equal or 
smaller than it is in the unmagnetized medium.  At $\Upsilon \leq 0.5$, the
effect of including the magnetic field on the drag force is
rather small. Interestingly,
at $\Upsilon \geq 0.5$, the strength of the drag 
for Mach numbers $> (1.7+\Upsilon^{2})^{1/2}$
is identical as it is in the unmagnetized case.

For $\Upsilon=1.41$, the drag force shows a plateau between
${\mathcal{M}}=0.6$ and $1.4$ (see Figure \ref{fig:fig_90g}). In general,
the drag force is remarkably supressed at Mach numbers around $\sim 1$
 in the
magnetized case as compared to the unmagnetized case, as long as
$\Upsilon>0.5$ (see, for instance the drag at $0.9\leq {\mathcal{M}}\leq 1.4$
for $\Upsilon=1$). In fact, the
temporal evolution of the drag force is given in Figure \ref{fig:force_time_90g}
for $\Upsilon=1$.  
The DF force on perturbers moving at Mach numbers $0.75\leq {\mathcal{M}}\leq 1.2$
saturates asymptotically to a constant value.
Hence, at $0.75\leq \mach< 1$, the drag forces reach a steady-state value either
the medium is magnetized or not.
However, we know that the unmagnetized drag force increases logarithmically 
in time for supersonic perturbers. This implies that 
the drag force may be suppressed by one order of magnitude at $1<{\mathcal{M}}<1.2$ in
the magnetized as compared to the unmagnetized medium because
the magnetized drag force saturates to a constant value.
On the other hand, at ${\mathcal{M}}<0.75$ (again $\Upsilon=1$)
there is no indication that the drag force saturates, at least up to $t=40t_{\rm cross}$,
implying that the DF force in a magnetized medium may be larger by a factor of a few 
than the drag in the unmagnetized case. 

In summary, when the magnetic fields
are relevant, that is for $\Upsilon>0.5$, we distinguish three ranges.
At high Mach numbers [i.e.~${\mathcal{M}}> (1.7+\Upsilon^{2})^{1/2}$], 
the drag is the same as in the unmagnetized case.
At intermediate Mach numbers
($1<{\mathcal{M}}\leq {\mathcal{M}}_{\rm crit}^{-1}$), the drag is highly suppressed. Finally,
at low Mach numbers, the drag force is stronger in the MHD case
than in a purely hydrodynamical medium.

\subsection{Intermediate angle between perturber's velocity and
magnetic field. $\Theta=45^{\circ}$}

The upper panels of Figure \ref{fig:fig_map_45+90} exhibit the complex
morphology of the wake when $\Theta=\pi/4$. As it is obvious from
these panels, the gravitational drag will have two components: one parallel
to $\VV_{0}$, which produces the drag and loss of kinetic energy by the
perturber, and one component perpendicular
to $\VV_{0}$, which would change the direction of $\VV_{0}$. 
Given the symmetry of the problem, both components lie
in the $(y,z)$-plane. We will start our discussion by considering the gravitational
drag force.

For those simulations presented in Figure \ref{fig:fig_45g} with $\Theta=\pi/4$, the drag force
saturates within $t=40t_{\rm cross}$ only for the model with ${\mathcal{M}}=0.9$
and $\Upsilon=0.71$. The maximum of the drag force occurs at
Mach numbers near $\simeq {\mathcal{M}}_{\rm crit}^{-1}$.
At intermediate Mach numbers, say at $1.2$--$1.6$ for $\Upsilon=1.41$,
the drag force may decrease by a factor of $2$--$3$ as compared to the
force without magnetic fields.
At high Mach numbers, ${\mathcal{M}}>{\mathcal{M}}_{\rm crit}^{-1}$,
the drag force is slightly suppressed as compared to the unmagnetized
case, but this reduction is more modest than for $\Theta=0^{\circ}$.
Once again, at low Mach numbers ($\mach<0.75$), the drag force is stronger than
in the unmagnetized case.

As already said, the component of the force perpendicular to the 
velocity of the perturber, $F_{\rm perp}$,
will tend to induce a change in the direction of the velocity 
(note that we force
the body to move along a straight line). We will use the following
sign convention for $F_{\rm perp}$. For an angle $\Theta$ in the
interval
$0\leq\Theta\leq\pi/2$, $F_{\rm perp}>0$ will mean that $\dot{\Theta}>0$, in our
convention.
In Figure \ref{fig:fig_45g}, $F_{\rm perp}$ is shown as a function 
of Mach number.  The magnitude of $F_{\rm perp}$
may be comparable to the drag force. For instance, at ${\mathcal{M}}=1.2$,
the perpendicular force is only a factor of $2$ smaller for $\Upsilon=1.41$
and a factor of $3$ for $\Upsilon=1$. Given a certain supersonic velocity, $F_{\rm perp}$ increases
with $\Upsilon$, while $F_{DF}$ shows the opposite behaviour.
For supersonic motions with angle $\Theta=\pi/4$,  $F_{\rm perp}$ is always positive
and increases monotonically in time (see Fig.~\ref{fig:fig_force_perp_45g}).
This implies that $F_{\rm perp}$ will tend to redirect perturber's velocity to a higher $\Theta$.
For the cases shown in Figure \ref{fig:fig_45g}, the perpendicular force
saturates in the run of the calculation ($t=40t_{\rm cross}$) only in two cases; for ${\mathcal{M}}=0.5$ and
$\Upsilon=0.71$, and for ${\mathcal{M}}=0.75$ and $\Upsilon=1.41$ (this latter case is
shown in Fig.~\ref{fig:fig_force_perp_45g}).
In some cases with subsonic Mach numbers, the perpendicular component of the
force is initially positive, achieves a maximum and then stars a linear decline up to
negative values (see Fig.~\ref{fig:fig_force_perp_45g}). 

\subsection{Dependence of the drag force on $\Theta$}
\begin{figure}
  \epsscale{0.77}
  \plotone{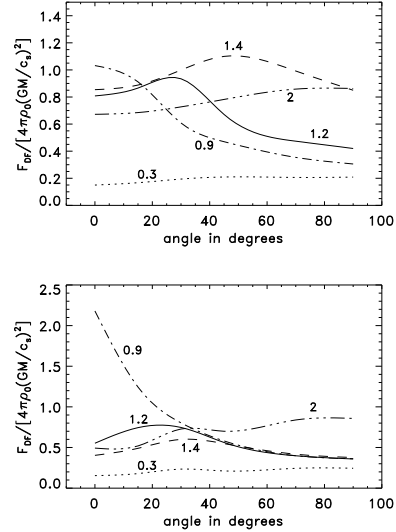}
  \caption{\label{fig:fig_force_angle}
Dependence of the drag force on the angle $\Theta$ for $\Upsilon=1$
(upper panel) and $\Upsilon=1.41$ (lower panel). The
numbers given at each curve represent the Mach number ${\mathcal{M}}$. 
The drag force was computed at $t=40t_{\rm cross}$.
}
\end{figure}

In Figure \ref{fig:fig_force_angle} we plot the drag force as a function of $\Theta$,
for $\Upsilon=1$ and $1.41$.  The dependence of $F_{DF}$ on $\Theta$ is not
always monotonic.
The strongest variation of $F_{DF}$ with $\Theta$ occurs for 
${\mathcal{M}}=0.9$. 
For this Mach number, the drag force may decrease by a factor of $2$--$3$
from $\Theta=0$ to $\Theta=30^{\circ}$.  For ${\mathcal{M}}=1.2$,
the drag force may change up to a factor of $2$ depending on the $\Theta$-value.
For ${\mathcal{M}}=0.3$ and $1.4$, the drag force depends gently 
on the angle.

In many astrophysical scenarios, the perturber will be subject to
an external gravitational potential and will describe a nonrectilinear
orbit. 
S\'anchez-Salcedo \& Brandenburg (2001) numerically treated the orbital
decay of a perturber in orbit around a unmagnetized gaseous sphere. 
They found that the ``local approximation'', that is estimating the drag force at the
present location of the perturber as if the medium were homogeneous but taking
appropriately the Coulomb logarithm, is very successful.
Consider now a perturber on a circular orbit in
a magnetized medium. If the orbit lies in
a plane perpendicular to the magnetic field, the attack angle $\Theta$ is
always $\pi/2$. In the local approximation, the maximum drag for
$\Theta=\pi/2$ and $\Upsilon=0$ occurs at $\mach\approx 1$ and at $\mach\approx 1.7$ for
$\Upsilon=1$. However, if the plane of the circular orbit is parallel to the
direction of the initial magnetic field, $\Theta$ will change periodically in time
as $\Theta= \Theta_{0}+\Omega_{0}t$.
Therefore, if the local approximation is valid, one can estimate the
mean drag force over a rotation period, which is approximately equivalent to take
the mean value of $F_{DF}$ over $\Theta$. In particular, for $\Upsilon=1$,
the $\Theta$-average drag force is maximum at $\mach\approx 1.4$.
This example illustrates how $F_{DF}$ may depend on $\Upsilon$ and on the
inclination of the orbit respect to the magnetic field lines. A more detailed analysis
of the drag force on a body on a circular orbit will be given somewhere else.

\section{Summary and discussion}
\label{sec:summary}
Understanding the nature of the DF force experienced 
by a gravitational object that moves against a mass
density background is of great importance to describe the evolution of gravitational
systems. In this work, we investigated the DF on a body moving in rectilinear trajectory
through a gaseous medium with a magnetic field uniform on the scales
considered.
In linear theory, the problem is largely characterized by three 
dimensionless parameters, $\mach$, which is defined as the ratio
of the particle velocity to the sound speed of the uniform gas, $\Upsilon$,
defined as the ratio between the Alfv\'en and sound speeds, and $\Theta$, the
angle between the magnetic field direction and the particle velocity. We find that magnetic 
effects may alter the drag force, especially for $\Upsilon>0.5$, 
because the magnetic field affects the flow velocity field 
in the perpendicular direction of the ambient field lines, and thereby
the morphology of the wake. Note that the plasma beta,
defined as the ratio of gas to magnetic 
pressure, is $2/\Upsilon^{2}$ for an isothermal system. 

There are two major differences between the magnetized and unmagnetized
case.
One conceptual difference is that, while gravitational focusing in a unmagnetized
medium always generates a positive density
enhancement, this is not the case in a magnetized medium
(see, e.g., Figs.~\ref{fig:map_0_141}, \ref{fig:map_07+09} and
\ref{fig:fig_map_45+90}). 
A second result is that
the peak value of the drag force is not near $\mach=1$ for a mass moving in a magnetized
medium. In fact,
the sharp peak of $F_{DF}$ at $\mach=1$ found in the $\Upsilon=0$ case
is no longer present in a magnetized medium with $\Upsilon>0.5$. For instance,
for a perturber in perpendicular motion to the field lines 
($\Theta=\pi/2$) in a medium with $\Upsilon=1.41$, 
the drag force is essentially constant from $\mach=0.5$ to $1.4$
and its maximum is located around $\mach=2$ (see Fig.~\ref{fig:fig_90g}).  
The flat plateau in the drag force between $\mach=0.5$ and $1.4$ is
partly because of the extra rigidity of the magnetic field in
the $x$ and $y$ directions.  

For a body traveling along the field lines, i.e.~$\Theta=0$, the steady-state problem can be treated analytically. We focus first on this case.
For $\Upsilon\neq 0$, the drag force presents two local maxima
(see Fig.~\ref{fig:dokuchaev}); one is located
in the subsonic branch 
(at $\mach_{\rm crit}$) and the other peak value is at the supersonic branch [at $\mach={\rm max}(1,\sqrt{2}\Upsilon)$]. 
When the velocity of the perturber is supersonic and super-Alfv\'enic
(and $\Theta=0$), the DF force in a magnetized medium is weaker than
it is in the unmagnetized case by a factor of $(1-\eta)$, 
with $\eta=(c_{A}/V_{0})^{2}$. 
The physical reason is that the medium becomes more rigid in the radial
direction and, hence, the opening aperture of the modified Mach cone is
the same as that in a unmagnetized medium with effective sound speed 
$(c_{s}^{2}+c_{A}^{2})^{1/2}$, but the density enhancement 
is smaller by a factor of $(1-\eta)$.
By contrast, the drag force for subsonic velocities is stronger if 
the medium is uniformly magnetized. For $\Theta=0$, an underdense
region is formed upstream because of the gas channeling
along the direction of the magnetic field, following the path
of less resistance.  
The steady-state theory predicts that the gravitational drag on a body with
$\Theta=0$ vanishes at Mach numbers in the
following two ranges: (1) at $\mach<\mach_{\rm crit}$ and (2) at ${\rm min}(1,\Upsilon)<\mach<
{\rm max}(1,\Upsilon)$. However, using time-dependent analysis we find that
the DF force asymptotically approaches to a nonzero steady-state value at these
Mach numbers.
For $\Upsilon>0.4$ (still $\Theta=0$), the drag force is maximized for
perturbers moving at a Mach number close to 
$\mach_{\rm crit}$ (Fig.~\ref{fig:fig_200_0g}). 
At Mach numbers around $\mach_{\rm crit}$, the density
enhancement is large but negative in a cone in front of the body. 
At those Mach numbers, the DF may be even 
more efficient than in the stellar case.  For example, for a medium with $\Upsilon=1.41$,
the drag force peaks between $\mach=0.6$ and $\mach=1.1$. As a consequence of the
stronger DF force, subsonic massive objects in a orbit elongated along the magnetic field lines
in a constant-density core of a nonsingular gaseous sphere will suffer a orbital decay faster
if the medium is pervased by a large-scale magnetic field.

We have also explored the $\Theta$-dependence of the DF drag.  
For Mach numbers around
$\mach_{\rm crit}$, the drag force exhibits the strongest variations 
with $\Theta$ (see Fig.~\ref{fig:fig_force_angle}).
For magnetized media with $\Upsilon \geq 1$ and
 regardless the exact value of $\Theta$, we find that
(1) the drag force for subsonic perturbers is higher by a factor of 
$1.5$--$2$ than it is for the unmagnetized DF drag,
and (2) for supersonic perturbers ($\mach>1$), the magnetized drag force is always weaker than the unmagnetized drag force.
At intermediate Mach numbers, $1.1\leq\mach\leq 1.4$, the drag force
is a factor of $2$--$3$ weaker than it is in the absence of magnetic fields\footnote{This factor
may be larger at later times because the magnetized drag force saturates,
whereas it increases logarithmically in time in the unmagnetized case. See Figure 
\ref{fig:fig_200_0g} for an evolved stage.}. 
At high Mach numbers, $\mach> (1.7+\Upsilon^{2})^{1/2}$, the suppresion of the drag
force is more important at small values of $\Theta$ 
(Fig.~\ref{fig:fig_force_angle}).
At these high Mach numbers and for an angle of 
$\Theta=\pi/2$, the drag forces are similar with and without 
magnetic fields (Fig.~\ref{fig:fig_90g}). 

As a consequence, supersonic massive objects
may make their way more slowly to the center of the system if the medium is pervased by
a large-scale magnetic field. As a model problem, consider a singular isothermal spherical
cloud threaded by a uniform magnetic field and a small-scale random magnetic field with
Alfv\'en speed $c_{a}$ everywhere constant. The density profile of the cloud is given
by $\rho(r)=(c_{s}^{2}+c_{a}^{2}/2)/2\pi G r^{2}$, where $c_{s}$ is the
isothermal sound speed.
The circular speed is $V_{0}=\sqrt{2c_{s}^{2}+c_{a}^{2}}$.  Since the effective sound
speed is $\sqrt{\gamma c_{s}^{2}+2c_{a}^{2}/3}$, the Mach number of a body on a quasi-circular
orbit is 
\begin{equation}
\mach=\left(\frac{2c_{s}^{2}+c_{a}^{2}}{\gamma c_{s}^{2}+\frac{2}{3}c_{a}^{2}}\right)^{1/2},
\end{equation}
which varies from
$1.1$ to $1.4$ depending on the value of $c_{a}$ and whether the 
perturbations are isothermal or adiabatic\footnote{In the nonmagnetic
simulations of the orbital decay of a single black
hole due to gaseous DF in Escala et al.~(2004), the velocity of the black hole is initially
supersonic ($\mach=1.4$) and remains barely supersonic through most of the
simulation.}.
If the uniform magnetic field component
has a $\Upsilon$-value between $1$ and $1.41$, the time for the perturber's orbit
to decay will be a factor of $2$--$5$ larger than the corresponding decay time for 
$\Upsilon=0$. Our results
demonstrate that,  in the presence of ordered magnetic fields with $\Upsilon>0.7$,
the role of the magnetic field on the drag force should be taken into account
to have accurate estimates of the timescales of orbital decay via gravitational DF.

\acknowledgements
The author would like to thank Miguel Alcubierre and Juan Carlos
Degollado for interesting discussions. Constructive comments
by an anomymous referee are greatly appreciated. This work has been partly
supported by CONACyT project 60526.

\appendix
\section{A. Fourier transformation: Axisymmetric case}
\label{sec:appendix1}
The three-dimensional Fourier transform 
of a perturbed variable $f(\rr)$ is given by
\begin{equation}
\hat{f}(\kk)=\frac{1}{(2\pi)^{3/2}}\int_{{\mathcal{R}}^{3}}
f(\rr) \, e^{-i\kk\cdot\rr} d^{3}\rr.
\end{equation}

In the Fourier space, Equations (\ref{eq:steady_ostriker_gen}) and (\ref{eq:goesFourier2}) are 
transformed into:
\begin{equation}
{\mathcal{M}}^{2} k_{z}^{2}\hat{\alpha}=k^{2} \hat{\alpha}-
\frac{4\pi G}{c_{s}^{2}} \hat{\rho}_{p}+\Upsilon^{2}
k^{2} \hat{\beta}_{z},
\label{eq:take1}
\end{equation}
\begin{equation}
i({\mathcal{M}}^{2}-1) k_{z}\hat{\alpha}=
\frac{i}{c_{s}^{2}} k_{z}\hat{\Phi}
+i{\mathcal{M}}^{2} k_{z}\hat{\beta}_{z},
\label{eq:take2}
\end{equation}
where $\rho_{p}$ is the mass density of the perturber, thus $\nabla^{2}\Phi=4\pi G\rho_{p}$.
In order to have an equation for $\hat{\alpha}$, 
we will eliminate $\hat{\beta}_{z}$. From Eq.~(\ref{eq:take2}), we have
\begin{equation}
\hat{\beta}_{z}=\frac{1}{{\mathcal{M}}^{2}} \left[({\mathcal{M}}^{2}-1)
\hat{\alpha}-\frac{\hat{\Phi}}{c_{s}^{2}}\right],
\end{equation}
and substituting into Eq.~(\ref{eq:take1}) we find
\begin{equation}
\left[{\mathcal{M}}^{2} k_{z}^{2}-k^{2}\left(1+
(1-\mach^{-2})\Upsilon^{2}\right)\right]\hat{\alpha}=
\frac{4\pi G}{c_{s}^{2}} 
\left(\frac{\Upsilon^{2}}{{\mathcal{M}}^{2}}-1\right) \hat{\rho}_{p},
\label{eq:fourier_parallel}.
\end{equation}

In the absence of magnetic fields ($\Upsilon=0$),
the above equation reduces to
\begin{equation}
({\mathcal{M}}^{2}k_{z}^{2}-k^{2})\hat{\alpha}
=-\frac{4\pi G}{c_{s}^{2}} \hat{\rho}_{p},
\label{eq:nofield}
\end{equation}
and the standard steady-state equation for the wake past a gravitating
body is recovered.  
At velocities much larger than the Alfv\'en speed, ${\mathcal{M}}\gg c_{A}/c_{s}=\Upsilon$,
Equation (\ref{eq:fourier_parallel}) is simplified to 
\begin{equation}
\left[\left(1+\Upsilon^{2}\right)^{-1}{\mathcal{M}}^{2}k_{z}^{2}-k^{2}\right]\hat{\alpha}
=-\frac{4\pi G}{(1+\Upsilon^{2})c_{s}^{2}} \hat{\rho}_{p}.
\label{eq:largeMach}
\end{equation}
By comparing the above equation with Equation (\ref{eq:nofield}), we see that the response of the
medium in this case is indistinguishable to that of an unmagnetized medium with sound speed 
$(c_{s}^{2}+c_{A}^{2})^{1/2}=c_{s}(1+\Upsilon^{2})^{1/2}$.

It is interesting to note that when $V_{0}=c_{A}\neq 0$, the right-hand-side
of Equation (\ref{eq:fourier_parallel}) vanishes and thereby the solution
is $\alpha=0$, implying that the steady-state configuration satisfies
$\nab\cdot\vv'=0$. Obviously, the drag force is exactly
zero in this configuration.

There exist two situations where the differential equation (\ref{eq:fourier_parallel}) is not well-posed:
(1) at ${\mathcal{M}}=1$ and (2) at the critical Mach number, $\mach_{\rm crit}$, satisfying that 
\begin{equation}
1+(1-\mach_{\rm crit}^{-2})\Upsilon^{2} =0.
\end{equation}
So that
\begin{equation}
{\mathcal{M}}_{\rm crit}\equiv \left(1+\Upsilon^{-2}\right)^{-1/2}.
\end{equation}
It is clear that ${\mathcal{M}}_{\rm crit}<1$.  If the dynamics is dominated by the magnetic field,
i.e. when $\Upsilon\gg 1$, then ${\mathcal{M}}_{\rm crit}\rightarrow 1$.
If not specified, we will consider ${\mathcal{M}}\neq 1$ and 
$\mach\neq {\mathcal{M}}_{\rm crit}$ throughout this section.

We will now calculate the solution of Eq.~(\ref{eq:fourier_parallel}) 
when the perturber is a point mass $M$,
so that $\hat{\rho}_{p}=M/(2\pi)^{3/2}$,
which corresponds to the Fourier transformation of 
$\rho_{p}(\rr)=M\delta(\rr)$,  
to obtain the Green's function. Using the convolution theorem, it is
possible to evaluate $\alpha$ for any general distribution $\rho_{p}$.
Hence, we solve for
\begin{equation}
\alpha(\rr) =\frac{(1-\eta)GM}{2\pi^{2}c_{s}^{2}}
\int \frac{1}{\xi k^{2}-{\mathcal{M}}^2 k_{z}^{2}}
e^{i\kk\cdot\rr} d^{3}\kk,
\label{eq:alphaks}
\end{equation}
where
\begin{equation}
\eta=\left(\frac{\Upsilon}{\mach}\right)^{2}=
\left(\frac{c_{A}}{V_{0}}\right)^{2},
\end{equation}
and
\begin{equation}
\xi= 1+(1-\mach^{-2})  \Upsilon^{2}.
\end{equation}
The integral (\ref{eq:alphaks}) along $k_{z}$ 
is evaluated by transforming to the complex plane.
It is convenient to define $\gamma^{2}\equiv 1-{\mathcal{M}}^{2}/\xi$.
Either if $\mach$ stands in the range $0<{\mathcal{M}}<{\mathcal{M}}_{\rm crit}$ or in the range
${\rm min}\left(1,\Upsilon\right)<{\mathcal{M}} <{\rm max}\left(1,\Upsilon\right)$,
then $\gamma^{2} >0$ and thus 
none of the poles lie on the real axis. Hence
we may use Cauchy's residue theorem to obtain:
\begin{equation}
\alpha(\rr)=\frac{(1-\eta)GM}{\xi c_{s}^{2}}
\frac{1}{\sqrt{z^{2}+R^{2}\gamma^{2}}}.
\end{equation}

For Mach numbers in any of the two ranges:
${\mathcal{M}}_{\rm crit}< {\mathcal{M}} < {\rm min}\left(1,\Upsilon\right)\,\,\,\,\,\,\,
{\rm and}\,\,\,\,\,\,\, {\mathcal{M}}>{\rm max}\left(1,\Upsilon\right)$,
the integrand has poles on the real axis. Hence we make `indentations' in the contour
at the position of the poles. We consider first Mach numbers larger
than $ {\rm max}(1,\Upsilon)$. Then,
for $z>0$, we close the contour at $+i\infty$, 
leaving the poles outside the contour to preserve causality, whereas for $z<0$, 
we consider a domain containing the lower
half-plane, that is where ${\rm Im}(k_{z})<0$, and the contour
slightly above the real axis, so that the two poles lie inside
the contour. 
More specifically, for $z<0$, the integration over $k_{z}$ can be evaluated as 
\begin{equation}
\int_{-\infty}^{\infty} \frac{e^{ik_{z}z}}{k^{2}-{\mathcal{M}}^2 \xi^{-1}k_{z}^{2}} dk_{z}=
\int_{-\infty}^{\infty} \frac{e^{ik_{z}z}}{k_{x}^{2}+k_{y}^{2}-\gamma_{1}^{2}k_{z}^{2}} dk_{z}=
-\frac{2\pi}{\gamma_{1} k_{R}} \sin \left(\frac{k_{R}z}{\gamma_{1}}\right),
\end{equation}
where $\gamma_{1}^{2} \equiv -\gamma^{2} = \xi^{-1}{\mathcal{M}}^2-1$ and 
$k_{R}^{2}=k_{x}^{2}+k_{y}^{2}$.  
The integration over $k_{x}$ and $k_{y}$ can be carried out
in polar coordinates ($k_{x}=k_{R}\cos \phi$ and $k_{y}=k_{R}\sin\phi$):
\begin{eqnarray}
-\frac{2\pi}{\gamma_{1}} \int_{0}^{\infty} \int_{0}^{2\pi} \sin\left(\frac{k_{R}z}{\gamma_{1}}\right)
e^{ik_{R}\cos\phi} d\phi \,\,dk_{R} \nonumber\\
=-\frac{4\pi^{2}}{\gamma_{1}} \int_{0}^{\infty} 
\sin\left(\frac{k_{R}z}{\gamma_{1}}\right) J_{0}(k_{R}R) \,\,dk_{R} \nonumber\\
=\left\{ \begin{array}{ll}
 \frac{4\pi^{2}}{\sqrt{z^{2}-\gamma_{1}^{2}R^{2}}}\;\;\; & \mbox{if $z<-\gamma_{1}R$};\\
0 & \mbox{otherwise.}\end{array} \right. 
\end{eqnarray}

At Mach numbers in the interval ${\mathcal{M}}_{\rm crit}< {\mathcal{M}}<{\rm min}(1,\Upsilon)$,
causality is perserved at $z>0$ if we consider a domain containing the upper 
half-plane
and the two poles lie inside the contour.
Putting all together, the solution is
\begin{equation}
\alpha(\rr)=\frac{\lambda(1-\eta)GM}{\xi c_{s}^{2}}
\frac{1}{\sqrt{z^{2}+R^{2}\gamma^{2}}},
\end{equation}
where
              \[ \lambda = \left\{ \begin{array}{ll}
         2 & \mbox{if ${\mathcal{M}}>{\rm max}(1,\Upsilon)$ and $z/R<-|\gamma|$};\\
         2 & \mbox{if ${\mathcal{M}}_{\rm crit}<{\mathcal{M}}<{\rm min}(1,\Upsilon)$ and 
              $z/R>|\gamma|$};\\
         1 & \mbox{if ${\mathcal{M}}<{\mathcal{M}}_{\rm crit}$ or if ${\rm min}(1,\Upsilon)
         <{\mathcal{M}}<{\rm max}(1,\Upsilon)$};\\
         0 & \mbox{otherwise.} \end{array} \right. \]
This result can be compared with that in Dokuchaev (1964) by noting that he used
the variable $A$ to denote the combination $(1-\eta)/\xi$.
Dokuchaev (1964) found the same functional form for $\alpha$, but failed to divide
correctly the cases according to the Mach number.
In the absence of a background magnetic field, which corresponds to $\xi=1$ and $\eta=0$,
we recover the classical form derived by previous authors.

\section{B. Fourier transformation: Perturber's velocity perpendicular 
to the magnetic field}
\label{sec:appendix2}

We consider a gravitational perturber moving at constant velocity
$v_{0} \hat{\yy}$ in an unperturbed medium with density $\rho_{0}$,
sound speed $c_{s}$ and a magnetic field $\BB_{0}=B_{0}\hat{\zz}$.
Therefore, the velocity of the perturber and the magnetic field are
perpendicular. We are interested in the stationary wake formed behind
the body. To do this, we assume that perturber is at rest and feels
a wind with velocity $-V_{0}\hat{\yy}$ at infinity.
Following the same approach as in Appendix \ref{sec:appendix1}, the two coupled governing
equations in this geometry are given by
\begin{equation}
{\mathcal{M}}^{2}\frac{\partial^{2} \alpha}{\partial y^{2}}
=\nabla^{2}\alpha +\frac{1}{c_{s}^{2}}\nabla^{2}\Phi
+\Upsilon^{2}\nabla^{2}\beta_{z},
\end{equation}
\begin{equation}
{\mathcal{M}}^{2}
\frac{\partial^{2} \alpha}{\partial y^{2}}=
\frac{\partial^{2} \alpha}{\partial z^{2}}
+\frac{1}{c_{s}^{2}}\frac{\partial^{2} \Phi}{\partial z^{2}}
+{\mathcal{M}}^{2}\frac{\partial^{2} \beta_{z}}{\partial y^{2}}.
\end{equation}
In the Fourier space, these equations are transformed into
\begin{equation}
-{\mathcal{M}}^{2} k_{y}^{2}\hat{\alpha}=-k^{2} \hat{\alpha}+
\frac{4\pi G}{c_{s}^{2}} \hat{\rho}_{p}-\Upsilon^{2}
k^{2} \hat{\beta}_{z},
\end{equation}
\begin{equation}
{\mathcal{M}}^{2} k_{y}^{2}\hat{\alpha}=k_{z}^{2}\hat{\alpha}
+\frac{1}{c_{s}^{2}} k_{z}^{2}\hat{\Phi}
+{\mathcal{M}}^{2} k_{y}^{2}\hat{\beta}_{z}.
\end{equation}
In order to have an equation for $\hat{\alpha}$, we eliminate $\hat{\beta}_{z}$
using
\begin{equation}
\hat{\beta}_{z}=
\frac{1}{\Upsilon^{2}k^{2}}\left[\frac{4\pi G}{c_{s}^{2}}\hat{\rho}_{p}
+({\mathcal{M}}^{2}k_{y}^{2}-k^{2})\hat{\alpha}\right],
\end{equation}
and we obtain the solution of the density disturbance in Fourier space:
\begin{eqnarray}
&&\left((1+\Upsilon^{2})k_{y}^{2}k^{2}
-{\mathcal{M}}^{2}
k_{y}^{4}
-\frac{\Upsilon^{2}}{{\mathcal{M}}^{2}}k_{z}^{2} k^{2}
\right) \hat{\alpha}=\nonumber\\
&&\frac{4\pi G}{c_{s}^{2}}\left( k_{y}^{2}-
\frac{\Upsilon^{2}}{{\mathcal{M}}^{2}}k_{z}^{2}\right)\hat{\rho}_{p}.
\end{eqnarray}
Of course, the structure of $\alpha$ in this case is  different than in the axisymmetric case  
(Eq.~\ref{eq:fourier_parallel}).
The inverse Fourier transform cannot be derived analytically but
since  the above equation is well-posed for 
any ${\mathcal{M}}$-value if $\Upsilon \neq 0$, we expect the drag force to be 
a continuous function of ${\mathcal{M}}$.


\begin{thebibliography}{}
\bibitem[Armitage \& Natarajan(2005)]{arm05}
  Armitage, P.\ J., \& Natarajan, P.\  2005, \apj, 634, 921
\bibitem[Bondi \& Hoyle(1944)]{bon44}
Bondi, H., \& Hoyle, F. 1944, \mnras, 104, 273
\bibitem[Bournaud et al.(2007)]{bou07}
Bournaud, F., Elmegreen, B. G., \& Elmegreen, D. M. 2007, \apj, 670, 237
\bibitem[Callegari et al.(2009)]{cal09}
Callegari, S., Mayer, L., Kazantzidis, S., Colpi, M., Governato, F.,
Quinn, T., \&  Wadsley, J. 2009, \apj, 696, L89
\bibitem[Cant\'o et al.(2011)]{can11}
Cant\'o, J., Raga, A. C., Esquivel, A., \& S\'anchez-Salcedo, F. J.
2011, \mnras, arXiv:1108.3032
\bibitem[Chandrasekhar(1943)]{cha43} Chandrasekhar, S.\ 1943, \apj, 97, 255
\bibitem[Chavarr\'{\i}a et al.(2010)]{cha10}
Chavarr\'{\i}a, L., Mardones, D., Garay, G., Escala, A., Bronfman, L., 
\& Lizano, S. 2010, \apj, 710, 583
\bibitem[Chuss et al.(2003)]{chu03}
Chuss, D. T., Davidson, J. A., Dotson, J. L., Dowell, C. D., 
Hildebrand, R. H., Novak, G., \& Vaillancourt, J. E. 2003, \apj, 599, 1116
\bibitem[Colpi \& Dotti(2011)]{col11}
Colpi, M., \& Dotti, M. 2011, Advanced Science Letters, 4, 181
\bibitem[Conroy \& Ostriker(2008)]{con08}
Conroy, C., \& Ostriker, J. P. 2008, \apj, 681, 151
\bibitem[Crocker et al.(2010)]{cro10}
Crocker, R. M., Jones, D. I., Melia, F., Ott, J., \& Protheroe, R. J.
2010, Nature, 463, 65
\bibitem[Dokuchaev(1964)]{dok64} Dokuchaev, V.~P.\ 1964, Soviet Astron., 8, 23
\bibitem[Dotti et al.(2006)]{dot06} Dotti, M., Colpi, M., \& Haardt, F.\ 2006, 
\mnras, 367, 103
\bibitem[El-Zant et al.(2004)]{elz04}
El-Zant, A. A., Kim, W.-T., \& Kamionkowski, M. 2004, \mnras, 354, 169
\bibitem[Escala et al.(2004)]{esc04}
Escala, A., Larson, R. B., Coppi, P. S., Mardones, D. 2004, \apj, 607, 765
\bibitem[Escala et al.(2005)]{esc05}
Escala, A., Larson, R. B., Coppi, P. S., Mardones, D. 2005, \apj, 630, 152
\bibitem[Goodman \& Heiles(1994)]{goo94}
Goodman, A. A., \& Heiles, C. 1994, \apj, 424, 208
\bibitem[Guedes et al.(2011)]{gue11}
Guedes, J., Madau, P., Mayer, L., \& Callegari, S. 2011, \apj, 729, 125
\bibitem[Heiles \& Crutcher(2005)]{hei05}
Heiles, C., \& Crutcher, R. 2005, Cosmic Magnetic Fields,
Notes in Physics, Eds. R. Wielebinski, R. Beck, 664, 137
\bibitem[Hornung, Pellat \& Barge(1985)]{hor85}
Hornung, P., Pellat, R., \& Barge, P. 1985, Icarus, 64, 295
\bibitem[Ida(1990)]{ida90}
Ida, S. 1990, Icarus, 88, 129
\bibitem[Immeli et al.(2004)]{imm04}
Immeli, A., Samland, M., Gerhard, O., \& Westera, P. 2004, \aap, 413, 547
\bibitem[Just \& Kegel(1990)]{jus90}
Just, A., \& Kegel, W. H. 1990, \aap, 232, 447
\bibitem[Kim(2007)]{kim07} Kim, W.-T.\ 2007, \apj, 667, L5
\bibitem[Kim(2010)]{kim10} Kim, W.-T.\ 2010, \apj, 725, 1069 
\bibitem[Kim et al.(2005)]{kim05} 
Kim, W.-T., El-Zant, A.~A., \& Kamionkowski, M.\ 2005, \apj, 632, 157
\bibitem[Kim \& Kim(2007)]{kim07b} Kim, H., \& Kim, W.-T.\ 2007, \apj, 665, 432 
\bibitem[Kim \& Kim(2009)]{kim09} Kim, H., \& Kim, W.-T.\ 2009, \apj, 703, 1278 
\bibitem[Kim, Kim \& S\'anchez-Salcedo(2008)]{kim08} 
Kim, H., Kim, W.-T., \& S\'anchez-Salcedo, F. J.\ 2008, \apj, 679, L33 
\bibitem[Lee \& Stahler(2011)]{lee11}
Lee, A. T., \& Stahler, S. W. 2011, \mnras, 416, 3177
\bibitem[Matthews \& Wilson(2002)]{mat02}
Matthews, B. C., \& Wilson, C. D. 2002, \apj, 574, 822
\bibitem[Maxted et al.(2009)]{max09}
Maxted, P. F. L. et al. 2009, \mnras, 400, 2012
\bibitem[Mayer et al.(2007)]{may07}
Mayer, L., Kazantzidis, S., Madau, P., Colpi, M., Quinn, T., \& Wadsley, J.
2007, Science, 316, 1874
\bibitem[Namouni(2010)]{nam10}
Namouni, F. 2010, \mnras, 401, 319
\bibitem[Namouni et al.(1996)]{nam96}
Namouni, F., Luciani, J. F., \& Pellat, R. 1996, \aap, 307, 972
\bibitem[Narayan(2000)]{nar00} Narayan, R.\ 2000, \apj, 536, 663
\bibitem[Nejad-Asghar(2010)]{nej10}
Nejad-Asghar, M. 2010, \mnras, 406, 1253
\bibitem[Nishiyama et al.(2010)]{nis10}
Nishiyama, S. et al. 2010, \apj, 722, L23 
\bibitem[Nordhaus \& Blackman(2006)]{nor06}
Nordhaus, J., \& Blackman, E. G. 2006, \mnras, 370, 2004
\bibitem[Ostriker(1999)]{ost99} Ostriker, E.~C.\ 1999, \apj, 513, 252
\bibitem[Rephaeli \& Salpeter(1980)]{rep80} Rephaeli, Y., \& Salpeter, E.~E.\ 1980, \apj, 240, 20
\bibitem[Ricker \& Taam(2008)]{ric08}
Ricker, P. M., \& Taam, R. E. 2008, \apj, 672, L41
\bibitem[Robishaw et al.(2008)]{rob08}
Robishaw, T., Quataert, E., \& Heiles, C. 2008, \apj, 680, 981
\bibitem[Ruderman \& Spiegel(1971)]{rud71} Ruderman, M.~A., \& Spiegel, E.~A.\ 1971, \apj, 165, 1
\bibitem[Ruffert(1996)]{ruf96}
Ruffert, M. 1996, \aap, 311, 817
\bibitem[S\'anchez-Salcedo(2009)]{san09}
S\'anchez-Salcedo, F. J. 2009, \mnras, 392, 1573
\bibitem[S{\'a}nchez-Salcedo \& Brandenburg(1999)]{san99} S{\'a}nchez-Salcedo, F.~J., \& Brandenburg, A.\ 1999, \apjl, 522, L35
\bibitem[S{\'a}nchez-Salcedo \& Brandenburg(2001)]{san01} S{\'a}nchez-Salcedo, F.~J., \& Brandenburg, A.\ 2001, \mnras, 322, 67
\bibitem[S\'anchez-Salcedo \& Santill\'an(2011)]{san11}
S\'anchez-Salcedo, F. J., \& Santill\'an, A. 2011, Rev.Mex.A.\&A., 47, 15
\bibitem[Sanders \& Mirabel(1996)]{san96}
Sanders, D. \& Mirabel, I. 1996, \araa, 34, 749
\bibitem[Stahler(2010)]{sta10}
Stahler, S. W. 2010, \mnras, 402, 1758
\bibitem[Steawart \& Wetherill(1988)]{ste88}
Stewart, G. R., \& Wetherill, G. W. 1988, Icarus, 74, 542
\bibitem[Taam \& Sandquist(2000)]{taa00}
Taam, R. E., \& Sandquist, E. L. \araa, 38, 113
\bibitem[Tamura \& Sato(1989)]{tam89}
Tamura, M., \& Sato, S. 1989, \aj, 98, 1368
\bibitem[Tanaka \& Haiman(2009)]{tan09}
Tanaka, T., \& Haiman, Z. 2009, \apj, 696, 1798
\bibitem[Thompson et al.(2009)]{tho09}
Thompson, T. A., Quataert, E., \& Murray, N. 2009, \mnras, 397, 1410
\bibitem[Thompson et al.(2006)]{tho06}
Thompson, T. A., Quataert, E., Waxman, E., Murray, N., \& Martin, C. L. 2006, \apj, 645, 186
\bibitem[Villaver \& Livio(2009)]{vil09}
Villaver, E., \& Livio, M. 2009, \apj, 705, L81
\bibitem[Yusef-Zadeh, Morris \& Chance(1984)]{yus84}
Yusef-Zadeh, F., Morris, M., \& Chance, D. 1984, Nature, 310, 557
\bibitem[Zweibel(2002)]{zwe02}
Zweibel, E. G. 2002, \apj, 567, 962
\end{thebibliography}
\end{document}